\documentclass[showpacs,twocolumn, preprintnumbers,amsmath,amssymb]{revtex4}
\usepackage{amsmath}
\usepackage{enumitem}
\usepackage{algorithm}
\usepackage{algpseudocode}
\usepackage{comment}
\usepackage{multirow}
\usepackage{algorithm}
\usepackage{adjustbox}
\usepackage{slashed}
\usepackage{bm}
\usepackage{array,booktabs}
\usepackage{color}
\usepackage{xcolor}
\def\be {\begin{equation}}
\def\ee {\end{equation}}
\def\bea {\begin{eqnarray}}
\def\eea {\end{eqnarray}}
\def\bnn {\begin{eqnarray*}}
\def\enn {\end{eqnarray*}}
\def\bf {\begin{figure}}
\def\ef {\end{figure}}
\def\bpm {\begin{pmatrix}}
\def\epm {\end{pmatrix}}
\def\bvm {\begin{vmatrix}}
\def\evm {\end{vmatrix}}

\def\ing {\includegraphics}
\def\tx {\textrm}
\def\tr {\tx{tr}}

\def\tw {\textwidth}
\def\nn {\nonumber}
\def\non {\nonumber\\}

\def\al {\left |}
\def\ar {\right |}
\def\tw {\textwidth}

\def\ds {\displaystyle}
\def\fs {\footnotesize}
\def\ns {\normalsize}

\def\f {\frac}

\def\ve {\varepsilon}
\def\et {\ve_{\tau}}
\def\t {\tau}
\def\i {\imath}
\def\dag {\dagger}
\def\gr {\nabla}

\def\k {\bm{k}}

\def\r {\bm{r}}
\def\c {\bm{c}}
\def\p {\bm{p}}
\def\q {\bm{q}}
\def\l {\bm{l}}
\def\P {\bm{\Phi}}
\def\A {\bm{A}}

\def\D {\mathcal{D}}
\def\N {\mathcal{N}}
\def\O {\mathcal{O}}
\def\Z {\mathcal{Z}}
\def\F {\mathcal{F}}
\def\M {\mathcal{M}}

\def\bg {\bm{\gamma}}

\makeatletter
\renewcommand*\env@matrix[1][\arraystretch]{
  \edef\arraystretch{#1}
  \hskip -\arraycolsep
  \let\@ifnextchar\new@ifnextchar
  \array{*\c@MaxMatrixCols c}}
\makeatother

\begin{document}
\title{Emergence of non-Fermi liquid type Weyl metals driven by doped magnetic impurities in spin-orbit coupled semiconductors}
\author{Kyoung-Min Kim$^{1}$, Jinsu Kim$^{2}$, Soo-Whan Kim$^{3}$, Myung-Hwa Jung$^{2}$, and Ki-Seok Kim$^{1}$}
\affiliation{$^{1}$Department of Physics, POSTECH, Pohang, Gyeongbuk 37673, Korea\\ $^{2}$Department of Physics, Sogang University, Seoul 04107, Korea \\ $^{3}$Max Plank POSTECH Center for Complex Phase Materials, Pohang, Gyeongbuk 37673, Korea}
\date{\today}

\begin{abstract}
Constructing an effective field theory in terms of doped magnetic impurities (described by an O(3) vector model with a random mass term), itinerant electrons of spin-orbit coupled semiconductors (given by a Dirac theory with a relatively large mass term), and effective interactions between doped magnetic ions and itinerant electrons (assumed by an effective Zeeman coupling term), we perform the perturbative renormalization group analysis in the one-loop level based on the dimensional regularization technique. As a result, we find that the mass renormalization in dynamics of itinerant electrons acquires negative feedback effects due to quantum fluctuations involved with the Zeeman coupling term, in contrast with that of the conventional problem of quantum electrodynamics, where such interaction effects enhance the fermion mass more rapidly. Recalling that the applied magnetic field decreases the band gap in the presence of spin-orbit coupling, this renormalization group analysis shows that the external magnetic field overcomes the renormalized band gap, allowed by doped magnetic impurities even without ferromagnetic ordering. In other words, the Weyl metal physics can be controlled by doping magnetic impurities into spin-orbit coupled semiconductors, even if the external magnetic field alone cannot realize the Weyl metal phase due to relatively large band gaps of semiconductors. Furthermore, we emphasize that quasiparticles do not exist in this emergent disordered Weyl metal phase due to correlations with strong magnetic fluctuations. This non-Fermi liquid type Weyl metal state may be regarded to be a novel metallic phase in the respect that a topologically nontrivial band structure appears in the vicinity of quantum criticality.
\end{abstract}

\maketitle

\section{Introduction} \label{sec:introduction}

The problem of doping magnetic impurities into semiconductors has been investigated for more than two decades, referred to as dilute magnetic semiconductors \cite{DMS_Review_I,DMS_Review_II,DMS_Review_III}. Several fundamental problems such as the nature of effective interactions between randomly doped magnetic impurities and the mechanism of spin polarization of itinerant electrons resulting from these random-positioned magnetic impurities had been discussed both extensively and intensively. Unfortunately, the final goal to achieve the ferromagnetic critical temperature of itinerant electrons at the order of room temperature has not been reached, yet.

In the present study we propose completely a novel aspect in the problem of dilute magnetic semiconductors: A Weyl metal phase arises as a result of such doped magnetic impurities in spin-orbit coupled semiconductors under external magnetic fields. It may not be completely a new idea that breaking time-reversal symmetry by either applying magnetic fields or doping magnetic ions turns spin-orbit coupled Dirac metals into Weyl metals \cite{Ref1,Ref2,Ref3}. An essential point is that applied magnetic fields of the order of $10$ $T$ are much smaller than the band gap of original semiconductor samples without magnetic impurities, given by the order of $10^2$ $meV$ $\sim$ $10^3$ $meV$. This implies that it is not possible to reach the Weyl metal phase only by applying external magnetic fields without doped magnetic impurities. Doping magnetic impurities such as $Eu$ and $Gd$ into spin-orbit coupled semiconductors, recent experiments could realize Weyl metal phases in $Eu_{x}Bi_{2-x}Se_{3}$ and $Gd_{x}Bi_{2-x}Te_{3-y}Se_{y}$, respectively, where the doping concentration covers from $2 \%$ to $4 \%$ approximately \cite{Experiments_Doping_MI}. Here, transport measurements have shown that negative magneto-resistivity appears only when the applied magnetic field is in parallel with the applied electrical current \cite{Experiments_Doping_MI}, referred to as the negative longitudinal magneto-resistivity and regarded to be a fingerprint of the Weyl metallic state \cite{WM_Boltzmann_KSK,Nonlinear_LMR_WM,WMD_KKM_16}.

We investigate the role of doped magnetic impurities in spin-orbit coupled semiconductors, where gapped itinerant electrons are described by a Dirac theory with a mass parameter. First, we consider the situation that magnetic impurities are randomly distributed in the vicinity of antiferromagnetic ordering \cite{FeBiTe_EXP_KSK}. This physical picture suggests an O(3) vector model with the relativistic dispersion for the dynamics of randomly distributed magnetic impurities, where the distribution function of the random mass term is set to be Gaussian with a zero average value and a finite variance. Second, we assume that the dominant interaction channel between doped magnetic impurities and gapped itinerant electrons is described by the Zeeman term, reformulated as a chiral-current minimal coupling term, where the O(3) vector field of the doped magnetic impurity plays the role of an emergent chiral gauge field in the dynamics of gapped itinerant electrons \cite{DMTS_KKM_15,Anomaly_Chiral_Current}.

Based on this effective field theory, we perform the perturbative renormalization group analysis up to the one-loop level. As a result, we find that the excitation gap for itinerant electrons of semiconductors acquires negative renormalization effects due to chiral current fluctuations driven by doped magnetic impurities. However, it turns out that the gap cannot be closed by such chiral-current fluctuations since their average vanishes in the vicinity of antiferromagnetic ordering. If ferromagnetic ordering is considered, the chiral-current flowing phase is realized, nothing but the Weyl metal state \cite{WMD_KKM_16,DMTS_KKM_15,Anomaly_Chiral_Current}. Recent experiments have shown antiferromagnetic ordering at low temperatures for $Eu_{x}Bi_{2-x}Se_{3}$ and $Gd_{x}Bi_{2-x}Te_{3-y}Se_{y}$ \cite{Experiments_Doping_MI}. Even if the gap cannot be closed by doped magnetic impurities alone, this situation opens new possibility: Applying external magnetic fields into these magnetically doped systems, renormalized band gaps would be closed to allow Weyl metal phases. Actually, our renormalization group analysis leads us to propose an interesting phase diagram in the plane of temperature and external magnetic field at a given disorder strength for magnetic impurities. A Weyl metal phase arises below a critical temperature and above a critical magnetic field, relatively low due to the role of doped magnetic impurities. In particular, we suggest a two-parameter scaling theory \cite{WMD_KKM_16} for the longitudinal magnetoconductivity near the semiconductor to Weyl-metal transition.

One interesting aspect of the emergent disordered Weyl metal state is that quasiparticles do not exist due to correlations with strong magnetic fluctuations. We claim that this non-Fermi liquid type Weyl metal state may be regarded to be a novel metallic phase in the respect that a topologically nontrivial band structure appears in the vicinity of quantum criticality.

The present manuscript is organized as follows. In Sec. \ref{sec:model} we construct an effective field theory for spin-orbit coupled semiconductors and doped magnetic impurities. In addition, we prepare for the renormalization group analysis, setting up the general structure of the renormalization group transformation. In Sec. \ref{sec:rganalysis} we perform the renormalization group analysis up to the one-loop level, based on the dimensional regularization. In Sec. \ref{sec:emergence-Weyl} we propose a phase diagram for our quantum phase transition from a spin-orbit coupled semiconducting phase to a Weyl metal phase, driven by doped magnetic impurities in the presence of external magnetic fields. In addition, we suggest a two-parameter scaling theory for the longitudinal magnetoconductivity near the semiconductor to Weyl-metal transition. In Sec. \ref{sec:discussion} we discuss various subjects on the emergence of non-Fermi liquid type Weyl metals due to strong fluctuations of localized magnetic moments: A. Model construction, B. Origin of the negative feedback effect on the mass gap, C. Role of the random-mass disorder, D. Role of potential scattering, E. Ward identity, and F. Higher order quantum corrections. In Appendixes we show all details of our perturbative renormalization group analysis.

\section{Model system} \label{sec:model}

\subsection{Effective field theory} \label{sec:action}

Our effective Hamiltonian consists of three main parts. The first describes the dynamics of electrons in topological or band insulators with strong spin-orbit coupling, given by a free Dirac theory as a minimal model,
\be\hat{H}_{f}=\sum_{\k}\psi^{\dag}_{\sigma a\k}\big(v\k\cdot\bm{\sigma}_{\sigma\sigma'}\otimes\tau^{z}_{ab}+mI_{\sigma\sigma'}\otimes\tau^{x}_{ab}\big)\psi_{\sigma'b\k},\ee
where the sign change of the mass parameter $m$ gives rise to a topological phase transition from a topological insulating state to a normal band insulating phase \cite{TI_Reviews}. Here, $a$ ($\sigma$) is a band (spin) index. An important parameter is the excitation gap $m$ of these electrons, given by the order of $10^2$ $meV$ $\sim$ $10^3$ $meV$, which cannot be closed by external magnetic fields alone as discussed before. This effective Dirac theory is proposed to describe itinerant electrons in $Bi_{2}Se_{3}$ ($Eu_{x}Bi_{2-x}Se_{3}$) and $Bi_{2}Te_{3-y}Se_{y}$ ($Gd_{x}Bi_{2-x}Te_{3-y}Se_{y}$) \cite{Experiments_Doping_MI}.

\bf[t]\centering\ing[width=0.4\tw]{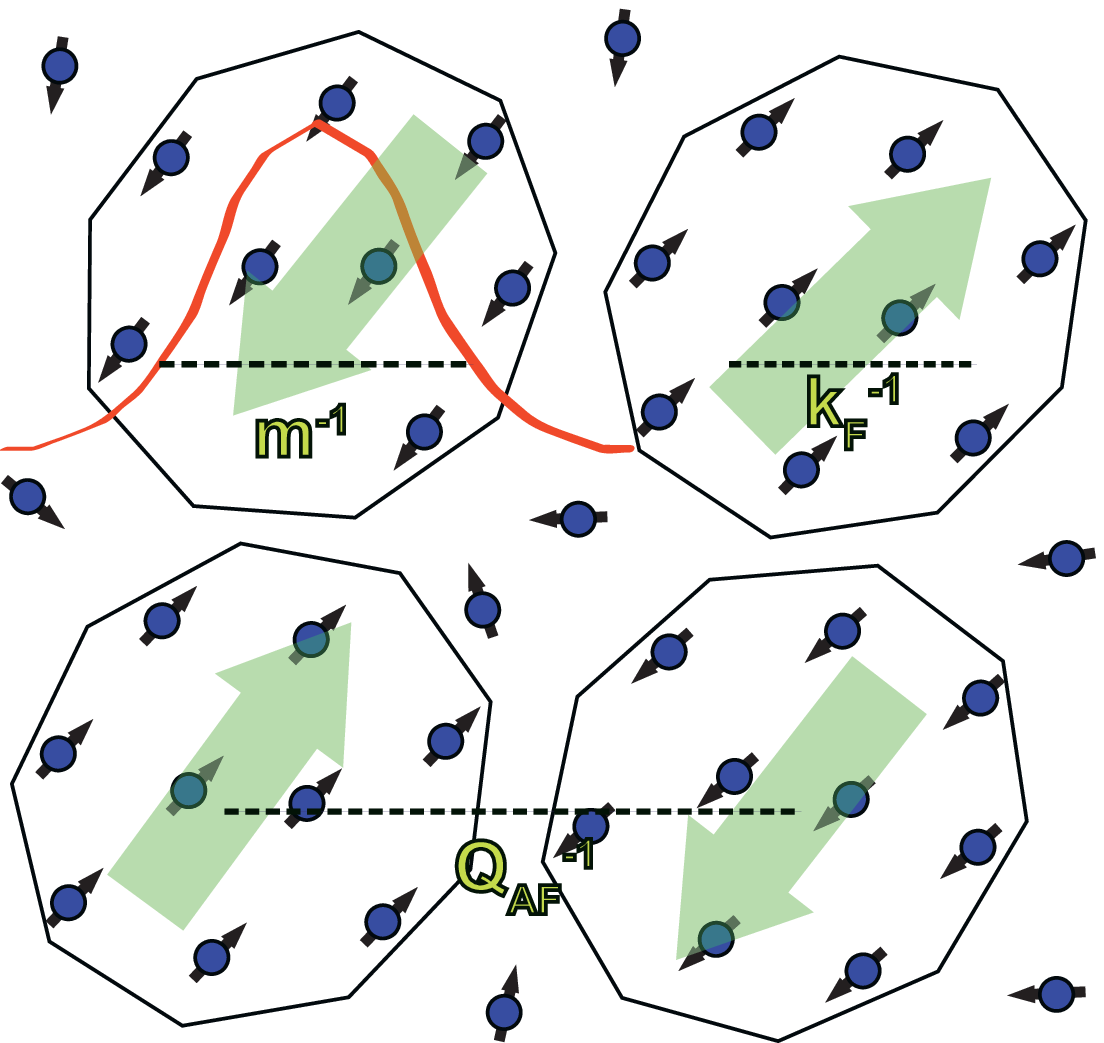}\caption{A schematic physical picture of a spin-orbit coupled semiconductor doped with magnetic impurities. Black arrows with blue balls denote doped magnetic impurity spins. Black polygons represent ferromagnetic clusters, correlated with each other via effective antiferromagnetic interactions. Red curve expresses the electron's wave function envelope.}\label{fig:picture}\ef

The second describes the dynamics of doped magnetic impurities. We propose an effective Hamiltonian for doped magnetic impurities as follows
\be\hat{H}_{m}=\sum_{ij}J_{ij}\bm{S}_{i}\cdot\bm{S}_{j},\ee
where $J_{ij}$ is a random variable, described by a probability functional of $P[J_{ij}]$. This effective Heisenberg model with random exchange interactions describes the dynamics of magnetic ions such as $Eu$ in $Eu_{x}Bi_{2-x}Se_{3}$ and $Gd$ in $Gd_{x}Bi_{2-x}Te_{3-y}Se_{y}$ \cite{Experiments_Doping_MI}. An essential question is how these magnetic impurities interact with each other in this almost insulating host, i.e., the nature of $J_{ij}$ and $P[J_{ij}]$ \cite{DMS_Review_I,DMS_Review_II,DMS_Review_III}. It turns out that samples in recent experiments are not in the insulating regime completely \cite{Experiments_Doping_MI,FeBiTe_EXP_KSK}. Instead, resistivity measurements show quite a small number of metallic carriers. Although this aspect does not mean that the Ruderman-Kittel-Kasuya-Yosida (RKKY) interaction would be the mechanism of effective interactions between doped magnetic moments \cite{RKKY}, where the dynamics of such electrons with a small Fermi surface would be essential to determine the nature of their effective interactions, we resort to this physical picture as our reference. The small Fermi surface leads us to consider ferromagnetic interactions dominantly between such magnetic moments, where the oscillating period of this effective interaction, given by the inverse of the Fermi momentum, is regarded to be quite long, compared with the average distance between magnetic impurities. As a result, it is natural to consider ferromagnetic clusters as coarse grained variables and their effective interactions. See Fig. \ref{fig:picture}. Measurements for spin susceptibility in recent experiments \cite{Experiments_Doping_MI,FeBiTe_EXP_KSK} show that an antiferromagnetic order appears around the order of $10$ $K$, depending on physical properties of magnetic impurities in samples, of course, including the concentration.

These experimental results drive us to construct an effective field theory for such ferromagnetic clusters $\bm{\Phi}$ in the form of an O(3) vector model with a relativistic dispersion relation \cite{FeBiTe_EXP_KSK}
\bea S_{m}&=&\int^{\beta}_{0}d\tau\int d^{3}\r\Big\{(\partial_{\tau}\bm{\Phi})^{2}+c^{2}(\partial_{\r}\bm{\Phi})^{2}\non
&&+(r+\delta r(\r))(\bm{\Phi}\cdot\bm{\Phi})+\f{u}{8}(\bm{\Phi}\cdot\bm{\Phi})^{2}\Big\}.\eea
Here, their random distributions in space are simulated by the introduction of a random mass term $\delta r(\r)$, regarded to be the most relevant term in this formulation. We recall that the transverse field Ising model can be mapped into an effective $\Phi^{4}$ theory in $(d+1)$ dimensions, assuming a paramagnetic ground state, where $d$ is a spatial dimension \cite{Phi4Theory_RG}. In other words, the transverse field Ising model shares essentially the same infrared universal physics with the $\Phi^{4}$ field theory, described by the Wilson-Fisher fixed point for quantum criticality \cite{Phi4Theory_RG}. Following the same procedure for the O(3) Heisenberg model, we obtain the above O(3) vector field theory. Actually, this mapping gives rise to other types of randomness in addition to the random mass term, given by random velocity and random self-interaction terms. Utilizing the replica trick and integrating the effective replica action over these random variables, one finds that these random variables are irrelevant at least for the weak disorder regime in the renormalization group analysis. In this respect we keep only the random mass term to describe the disordered nature of doped magnetic ions.

The most important third part is how these ferromagnetic clusters interact with electrons of the semiconducting host. Although it is not easy to determine the average size of such clusters, we propose an effective Zeeman interaction term between doped magnetic impurities and itinerant electrons, given by
\be\hat{H}_{int}=-\lambda\sum_{i}\psi_{i\sigma a}^{\dag}\bm{\sigma}_{\sigma\sigma'}\otimes\bm{I}_{ab}\psi_{i\sigma' b}\cdot\bm{S}_{i}.\ee
This effective interaction term is quite special in the respect that the chiral matrix appears in the mathematical expression, where the ferromagnetic cluster field drives the chiral current along all directions of fluctuations \cite{DMTS_KKM_15,Anomaly_Chiral_Current}. The mathematical formulation implies the conservation law of the U(1) chiral current. Indeed, the Ward identity is confirmed in the renormalization group analysis of the one-loop level, proven in Sec. \ref{sec:ward identity}.

Based on the above discussion, we construct an effective free energy functional as follows
\bea\F&=&-\frac{1}{\beta} \int\D\delta r(\r)P[\delta r(\r)]\ln\int\D \psi \D\bm{\Phi}e^{-S_{f}-S_{m}-S_{int}},\nn\\
S_{f}&=&\int^{\beta}_{0}d\tau\int d^{3}\r\bar{\psi}\big(\gamma_{0}\partial_{\tau}-\i v\bg\cdot\partial_{\r}+m\big)\psi,\nn\\
S_{m}&=&\int^{\beta}_{0}d\tau\int d^{3}\r\Big\{(\partial_{\tau}\bm{\Phi})^{2}+c^{2}(\partial_{\r}\bm{\Phi})^{2}\non
&&~~~~~~~~~~~~~~~~~+(r+\delta r(\r))(\bm{\Phi}\cdot\bm{\Phi})+\f{u}{8}(\bm{\Phi}\cdot\bm{\Phi})^{2}\Big\},\nn\\
S_{int}&=&-\int^{\beta}_{0}d\tau\int d^{3}\r\lambda\bar{\psi}\bm{\gamma}\gamma_{5}\psi\cdot\bm{\Phi}. \label{EFT_Free_Energy} \eea
Here, the Dirac theory is reformulated in terms of Dirac gamma matrices, given by \be \gamma_{0}=\bpm0&1\\1&0\epm,~\gamma_{i}=\bpm0&-\sigma_{i}\\\sigma_{i}&0\epm,~\gamma_{5}=\bpm1&0\\0&-1\epm \ee with ($i=1,~2,~3$), where $\psi=(\psi_{\uparrow1}~\psi_{\downarrow1}~\psi_{\uparrow2}~\psi_{\downarrow2})^{T}$ is a Dirac spinor field. In particular, we point out that the Zeeman interaction term is rewritten in the form of the chiral-current and gauge-field minimal coupling term, where the three-component vector field of $\bm{\Phi}$ plays the role of the chiral gauge field. We assume the Gaussian distribution for the random mass, given by $P[\delta r(\r)]=\N\int\D\delta r(\r)\exp[-\int\f{d^{3}\r[\delta r(\r)]^{2}}{2\Gamma_{m}}]$, where $\N$ is the normalization constant and $\Gamma_{m}$ is the variance.

Resorting to the replica trick for the disorder average \cite{WMD_KKM_16,DMTS_KKM_15}, we reformulate our effective field theory as follows
\bea\Z&=&\int\D(\bar{\psi}^{(a)},\psi^{(a)})\D\bm{\Phi}^{(a)}e^{-S_{f}-S_{m}-S_{int}},\non
S_{f}&=&\int^{\beta}_{0}d\tau\int d^{3}\r\bar{\psi}^{(a)}\big(\gamma_{0}\partial_{\tau}-\i v\bg\cdot\partial_{\r}+m\big)\psi^{(a)},\non
S_{m}&=&\int^{\beta}_{0}d\tau\int d^{3}\r\Big\{(\partial_{\tau}\bm{\Phi}^{(a)})^{2}+c^{2}(\partial_{\r}\bm{\Phi}^{(a)})^{2}\non
&&~~~~~~~~~~~~~~~~+r(\bm{\Phi}^{(a)})^{2}+\f{u}{8}(\bm{\Phi}^{(a)}\cdot\bm{\Phi}^{(a)})^{2}\non
&&~~~~-\int^{\beta}_{0}d\tau'\f{\Gamma_{m}}{8}(\bm{\Phi}^{(a)}_{\tau}\cdot\bm{\Phi}^{(a)}_{\tau})(\bm{\Phi}^{(a')}_{\tau'}\cdot\bm{\Phi}^{(a')}_{\tau'})\Big\},\non
S_{int}&=&-\int^{\beta}_{0}d\tau\int d^{3}\r\lambda\bar{\psi}^{(a)}\bg\gamma_{5}\psi^{(a)}\cdot\bm{\Phi}^{(a)},\label{eq:action}\eea
where $(a)$ is the replica index. We recall that fermions ($\psi^{(a)}$) represent semiconductor bands while bosons ($\bm{\Phi}^{(a)}$) denote spin fluctuations of ferromagnetic clusters around antiferromagnetic ordering. They are characterized by the fermion velocity ($v$), the fermion mass ($m$), the boson velocity ($c$), and the boson mass ($r$). Bosons are self-interacting with their interaction strength ($u$), and disorder scattering with the disorder strength ($\Gamma_{m}$). Both fermions and bosons are interacting through the Zeeman coupling term with the interaction strength ($\lambda$). Total seven parameters define this effective field theory completely.

\subsection{Setup for renormalization group analysis} \label{sec:setup}

\subsubsection{Renormalized effective field theory within the dimensional regularization scheme}

We take the double $\ve$-expansion scheme \cite{double_epsilon}, where all coupling constants of the self-interaction strength, the disorder strength, and the Zeeman interaction strength can be treated as perturbations. We generalize not only the space dimension ($3\rightarrow d$) but also the time dimension ($1\rightarrow d_{\tau}$). Then, the bare effective action is given by
\bea S_{f}&=&\int d^{d_{\tau}}\bm{\t}\int d^{d}\r\bar{\psi}^{(a)}\big(\bg_{\tau}\cdot\partial_{\bm{\t}}-\i\bg\cdot\partial_{\r}+m\big)\psi^{(a)},\non\
S_{m}&=&\int d^{d_{\tau}}\bm{\t}\int d^{d}\r\Big\{(\partial_{\bm{\t}}\bm{\Phi}^{(a)})^{2}+c^{2}(\partial_{\r}\bm{\Phi}^{(a)})^{2}\non
&&~~~~~~~~~~~~~~~~~~~+r(\bm{\Phi}^{(a)})^{2}+\f{u}{8}(\bm{\Phi}^{(a)}\cdot\bm{\Phi}^{(a)})^{2},\non
&&~~~~-\int d^{d_{\tau}}\bm{\t}'\f{\Gamma_{m}}{8}(\bm{\Phi}^{(a)}_{\bm{\tau}}\cdot\bm{\Phi}^{(a)}_{\bm{\tau}})(\bm{\Phi}^{(a')}_{\bm{\tau}'}\cdot\bm{\Phi}^{(a')}_{\bm{\tau}'})\Big\},\non
S_{int}&=&-\int d^{d_{\tau}}\bm{\t}\int d^{d}\r\lambda\bar{\psi}^{(a)}\bg\gamma_{5}\psi^{(a)}\cdot\bm{\Phi}^{(a)},\eea
where $\bg_{\tau}=(\gamma_{0},\cdots,\gamma_{(d_{\tau}-1)})$ and $\bg=(\gamma_{1},\cdots,\gamma_{d})$ follow the Clifford algebra as $\{\gamma_{\tau i},\gamma_{j}\}=0$, $\{\gamma_{\tau i},\gamma_{\tau j}\}=2\delta_{ij}1_{2\times2}$, and $\{\gamma_{i},\gamma_{j}\}=-2\delta_{ij}1_{2\times2}$.

Dimensional analysis gives scaling dimensions of $[\psi]=\f{d+d_{\tau}-1}{2}$, $[\bm{\Phi}]=\f{d+d_{\tau}-2}{2}$, $[c]=0$, $[m]=1$, and
\be[u]=4-d-d_{\tau},~[\Gamma_{m}]=4-d,~[\lambda]=\f{4-d-d_{\tau}}{2}.\ee
This leads us to set $d_{\tau}=\ve_{\tau}$ and $d=4-\ve-\ve_{\tau}$ in the loop calculation so that all coupling constants are put into the perturbative regime, where $[u]=\ve$, $[\Gamma_{m}]=\ve+\ve_{\tau}$, and $[\lambda]=\f{\ve}{2}$. In principle, we set $\ve_{\tau} = 1$ and $\ve = 0$ in the last stage. Here, the fermion velocity $v$ is set to be unity. As a result, we have total six parameters to characterize our effective field theory.

Introducing all counterterms to cancel ultraviolet (UV) divergences from quantum fluctuations into the above bare action, we have an effective renormalized action
\bea S_{f}&=&\int d^{d_{\tau}}\bm{\tau}\int d^{d}\r\bar{\psi}^{(a)}_{r}\big(Z_{0}\bg_{\tau}\cdot\partial_{\bm{\tau}}-Z_{1}\i\bg\cdot\partial_{\r}\non
&&~~~~~~~~~~~~~~~~~~~~~~~~~~~~~~~~~~~~~~~+\mu Z_{m}m_{r}\big)\psi^{(a)}_{r},\non
S_{m}&=&\int d^{d_{\tau}}\bm{\tau}\int d^{d}\r\Big\{Z_{2}(\partial_{\bm{\tau}}\bm{\Phi}^{(a)}_{r})^{2}+Z_{c}c_{r}^{2}(\partial_{\r}\bm{\Phi}^{(a)}_{r})^{2}\non
&&+\mu^{2}Z_{r}r_{r}(\bm{\Phi}_{r}^{(a)})^{2}+\f{\mu^{\ve}Z_{u}u_{r}}{8}(\bm{\Phi}^{(a)}_{r}\cdot\bm{\Phi}^{(a)}_{r})^{2}\non
&&-\int_{\bm{\tau}'}\f{\mu^{\ve+\ve_{\tau}}Z_{\Gamma_{m}}\Gamma_{mr}}{8}(\bm{\Phi}^{(a)}_{\bm{\tau}r}\cdot\bm{\Phi}^{(a)}_{\bm{\tau}r})(\bm{\Phi}^{(a')}_{\bm{\tau}'r}\cdot\bm{\Phi}^{(a')}_{\bm{\tau}'r})\Big\},\non
S_{int}&=&-\int d^{d_{\tau}}\bm{\tau}\int d^{d}\r Z_{\lambda}\mu^{\ve/2}\lambda_{r}\bar{\psi}^{(a)}_{r}\bg\gamma_{5}\psi^{(a)}_{r}\cdot\bm{\Phi}^{(a)}_{r}.\eea
Here, $\mu$ is a renormalization scale ($[\mu]=1$). Field renormalization factors are introduced to relate bare fields with renormalized ones as follows: $\psi=Z_{\psi}^{1/2}\psi_{r}$ and $\bm{\Phi}=Z_{\Phi}^{1/2}\bm{\Phi}_{r}$ with $Z_{\psi}=Z_{1}(Z_{0}/Z_{1})^{d_{\tau}}$ and $Z_{\Phi}=Z_{2}(Z_{0}/Z_{1})^{d_{\tau}-2}$, where the subscript $r$ means ``renormalized". All other renormalized parameters are given by
\bea&&m_{r}=\mu^{-1}(Z_{1}/Z_{m})m,\non[1.5pt]
&&c_{r}=(Z_{2}/Z_{c})^{1/2}(Z_{0}/Z_{1})^{-1}c,\non[1.5pt]
&&r_{r}=\mu^{-2}(Z_{2}/Z_{r})(Z_{0}/Z_{1})^{-2}r,\non[1.5pt]
&&\lambda_{r}=\mu^{-\f{\ve}{2}}Z_{2}^{1/2}(Z_{0}/Z_{1})^{-1+\epsilon_{\tau}/2}\lambda\non[1.5pt]
&&u_{r}=\mu^{-\ve}(Z_{2}^{2}/Z_{u})(Z_{0}/Z_{1})^{-4+\epsilon_{\tau}}u,\non[1.5pt]
&&\Gamma_{mr}=\mu^{-\ve-\ve_{\tau}}(Z_{2}^{2}/Z_{\Gamma_{m}})(Z_{0}/Z_{1})^{-4}\Gamma_{m},\eea
where the Ward identity of $Z_{1}=Z_{\lambda}$ has been used.

Counterterms are given by singular quantum corrections
\bea&&\delta_{0}\gamma_{0}=-\i\partial_{k_{0}}\Sigma(k),~\delta_{1}\bm{\gamma}=\partial_{\k}\Sigma(k),~\delta_{m}m=\Sigma(k)|_{k=0},\non[1pt]
&&\delta_{2}=\partial_{q_{0}^{2}}\Pi(q),~\delta_{c}=\partial_{\q^{2}}\Pi(q),~\delta_{r}r=\Pi(q)|_{q=0},\non[1pt]
&&\delta_{\lambda}=-\lambda^{-1}\sum_{i}\delta\lambda(i),~\delta_{u}=u^{-1}\sum_{i}\delta u(i),\non
&&\delta_{\Gamma_{m}}=-\Gamma_{m}^{-1}\sum_{i}\delta\Gamma_{m}(i),\label{eq:counter}\eea
where all renormalization factors are related to these counterterms as $Z_{0}=1+\delta_{0}$, $Z_{1}=1+\delta_{1}$, $Z_{m}=1+\delta_{m}$, $Z_{2}=1+\delta_{2}$, $Z_{c}=1+\delta_{c}$, $Z_{r}=1+\delta_{r}$, $Z_{u}=1+\delta_{u}$, $Z_{\Gamma_{m}}=1+\delta_{\Gamma_{m}}$, and $Z_{\lambda}=1+\delta_{\lambda}$. Here, we resort to the minimal subtraction scheme and use the convention of $G^{-1}(k)=G^{-1}_{0}(k)-\Sigma(k)$ and $D^{-1}(k)=D^{-1}_{0}(q)-\Pi(q)$. $\Sigma(k)$ ($\Pi(q)$) are singular self-energy corrections for fermions (bosons), $\delta\lambda(i)$ are singular fermion-boson vertex corrections, $\delta u(i)$ are singular boson self-interaction vertex corrections, and $\delta\Gamma_{m}(i)$ are singular boson disorder-scattering vertex corrections. The meaning of $i$ will be clarified below, used to identify various Feynman diagrams.

\subsubsection{Renormalization group equations}

The renormalized Green's function is defined as
\bea&&\left\langle\bar{\psi}_{r}(k_{1})\cdots\psi_{r}(k_{m+1})\cdots\bm{\Phi}_{r}(k_{2m+1})\cdots\right\rangle\non
&&=G^{(m,n)}(\{k_{i}\};\bm{F},\mu)\delta^{d+d_{\tau}}\bigg(\sum_{i=1}^{2m}k_{i}-\sum_{j=2m+1}^{2m+n}k_{j}\bigg),~~~~\label{eq:green-def}\eea
where the coupling constants are put into a vector form of $\bm{F}=(\lambda,u,\Gamma_{m},m,c,r)$. This is related to the bare Green's function as
\bea&&G^{(m,n)}(\{k_{i}\};\bm{F},\mu)\non
&&~~~~=Z_{\psi}^{-m}Z_{\Phi}^{-\f{n}{2}}(Z_{0}/Z_{1})^{-\epsilon_{\tau}}G_{B}^{(m,n)}(\{k_{Bi}\};\bm{F}_{B}).~~~~\label{eq:green-relation}\eea

It is straightforward to show that the renormalized Green's function satisfies the following differential equation, referred to as the Callan-Symanzik equation for the Green's function,
\bea&&\bigg\{\sum_{i=1}^{2m+n}\bigg(z\k_{\tau,i}\cdot\gr_{\k_{\tau,i}}+\k_{i}\cdot\gr_{\k_{i}}\bigg)-\bm{\beta}\cdot\gr_{\bm{F}}\non
&&-2m\bigg(-\f{5-\ve}{2}+\eta_{\psi}\bigg)-n\bigg(-\f{6-\ve}{2}+\eta_{\phi}\bigg)\non
&&-\epsilon_{\tau}(z-1)-(4-\ve)\bigg\}G^{(m,n)}(\{k_{i}\};\bm{F},\mu)=0.\label{eq:callan-symanzik}\eea
Here, beta functions are expressed in terms of renormalization factors as follows
\bea\beta_{\lambda}&=&\lambda\bigg[-\f{\ve}{2}-\bigg(1-\f{\ve_{\tau}}{2}\bigg)(z-1)+\f{1}{2}\f{\partial\ln{Z_{2}}}{\partial\ln\mu}\bigg],\non
\beta_{u}&=&u\bigg[-\ve-(4-\ve_{\tau})(z-1)+\f{\partial\ln(Z_{2}^{2}/Z_{u})}{\partial\ln\mu}\bigg],\non
\beta_{\Gamma_{m}}&=&\Gamma_{m}\bigg[-\ve-\ve_{\tau}-4(z-1)+\f{\partial\ln(Z_{2}^{2}/Z_{\Gamma_{m}})}{\partial\ln\mu}\bigg],\non
\beta_{c}&=&c\bigg[-(z-1)+\f{1}{2}\f{\partial\ln{(Z_{2}/Z_{c})}}{\partial\ln\mu}\bigg],\non
\beta_{m}&=&m\bigg[-1+\f{\partial\ln(Z_{1}/Z_{m})}{\partial\ln\mu}\bigg],\non
\beta_{r}&=&r\bigg[-2-2(z-1)+\f{\partial\ln(Z_{2}/Z_{r})}{\partial\ln\mu}\bigg],\label{eq:beta functions}\eea
where we defined $\beta_{\lambda}\equiv\f{\partial\lambda}{\partial\ln\mu}$, $\beta_{u}\equiv\f{\partial u}{\partial\ln\mu}$, $\beta_{\Gamma_{m}}\equiv\f{\partial \Gamma_{m}}{\partial\ln\mu}$, $\beta_{c}\equiv\f{\partial c}{\partial\ln\mu}$, and $\beta_{r}\equiv\f{\partial r}{\partial\ln\mu}$. $z$ is the dynamical critical exponent, introduced to incorporate the space-time anisotropy. $\eta_{\psi}$ ($\eta_{\Phi}$) is the anomalous scaling dimension for the fermion (boson) field, describing its fractal behavior. They are given by
\bea&&z=1+\f{\partial\ln(Z_{0}/Z_{1})}{\partial\ln\mu},~\eta_{\psi}=\f{1}{2}\f{\partial\ln Z_{1}}{\partial\ln\mu}+\f{\ve_{\tau}}{2}(z-1),\non
&&~~~~~~~~~~\eta_{\Phi}=\f{1}{2}\f{\partial\ln Z_{2}}{\partial\ln\mu}+\bigg(\f{\ve_{\tau}}{2}-1\bigg)(z-1).\label{eq:anomalousdimensions}\eea

Solving the Callan-Symanzik equation at the fixed point, given by the fact that all beta functions vanish, we obtain the scaling expressions for both Green's functions of fermions and bosons, respectively,
\bea G(\k_{\tau},\k)&=&\f{1}{|\k|^{1-\ve_{\tau}(z-1)-2\eta_{\psi}}}\tilde{g}(|\k_{\tau}|^{1/z}/|\k|),\nn\\
D(\q_{\tau},\q)&=&\f{1}{|\q|^{2-\ve_{\tau}(z-1)-2\eta_{\Phi}}}\tilde{d}(|\q_{\tau}|^{1/z}/|\q|).\label{eq:green}\eea
Here, $\tilde{g}(|\k_{\tau}|^{1/z}/|\k|)$ ($\tilde{d}(|\q_{\tau}|^{1/z}/|\q|)$) is the scaling function of the fermion (boson) propagator, which should be found by explicit calculations, not trivial.

\section{Renormalization group analysis} \label{sec:rganalysis}

\subsection{Self-energy corrections} \label{sec:self energy}

\bf[t]\centering\ing[width=0.45\tw]{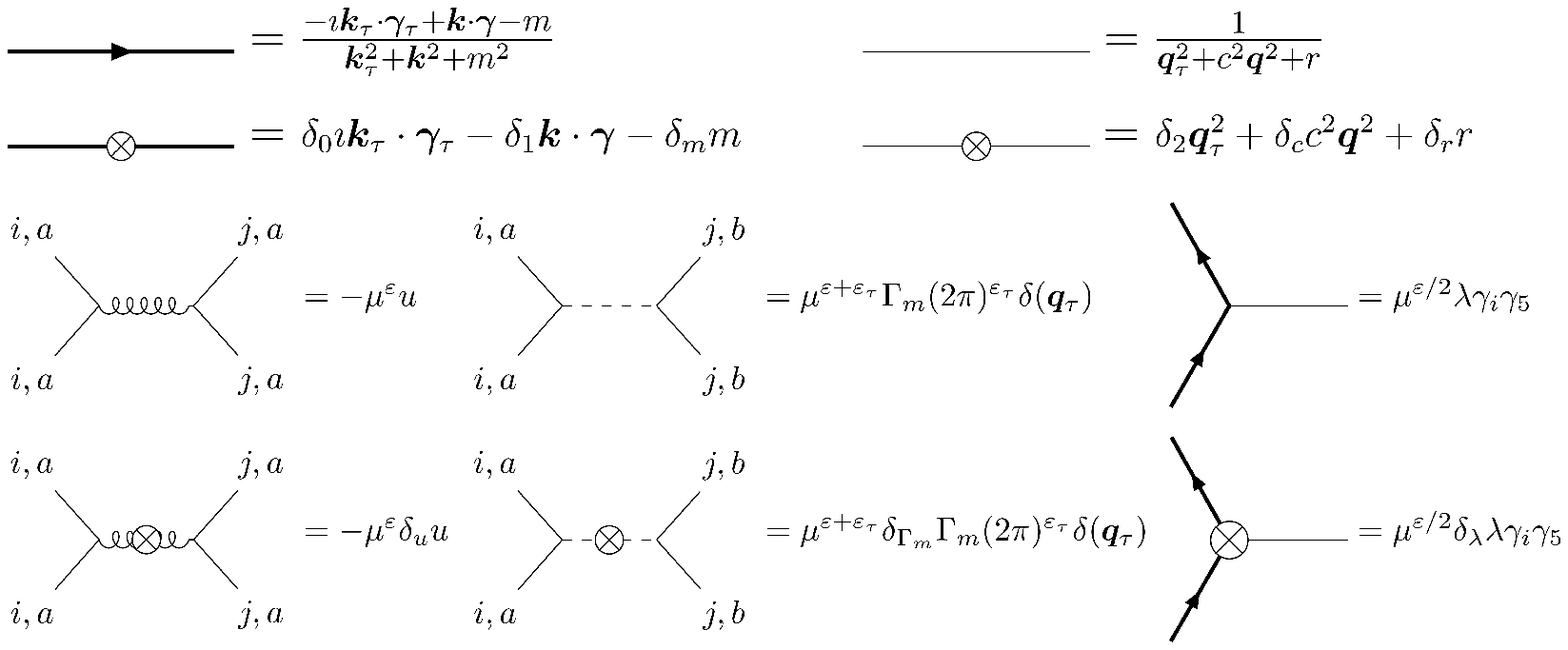}\caption{Feynman rules in momentum space. The first line represents fermion and boson propagators, respectively. The second line expresses counterterms for fermion and boson propagators, respectively. The third line describes three types of vertices for boson self-interactions, boson disorder scattering, and fermion-boson Zeeman interactions, respectively. The last line denotes counterterms for all interaction vertices described by the third line.}\label{fig:feynman rules}\ef

Based on the renormalized effective action, we introduce Feynman rules as shown in Fig. \ref{fig:feynman rules}. Here, the thick line represents an electron propagator, and the thin line describes the Green's function of an order-parameter field. The spring line means an effective self-interaction between order parameter fluctuations, and the dotted line gives an interaction vertex involved with disorder. The $\bigotimes$-symbol describes a counterterm for each propagator and each vertex. Resorting to these Feynman rules, one can take into account quantum fluctuations perturbatively, where the dimensional regularization scheme is utilized.

\bf[t]\centering\ing[width=0.4\tw]{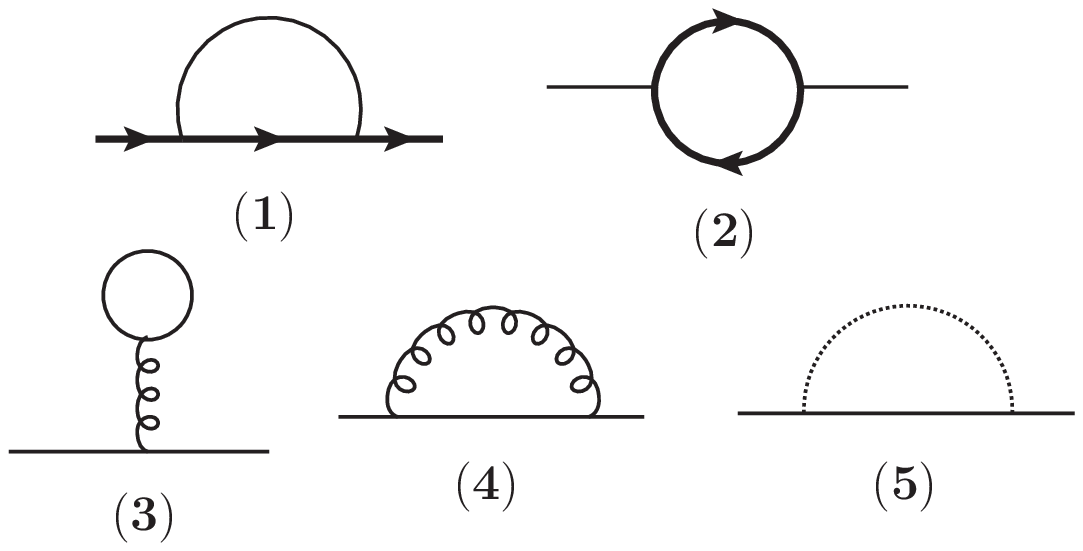}\caption{Self-energy corrections for both fermions and bosons. The first diagram is the fermion's self-energy and the others are the boson's self-energy corrections.} \label{fig:self-energy} \ef

The fermion self-energy is shown in Fig. \ref{fig:self-energy}-(1). The calculation is similar to that of quantum electrodynamics (QED) \cite{Phi4Theory_RG}. Here, we summarize our results
\bea&&Z_{0}-1=-\f{3\lambda^{2}}{4\pi^{2}\ve c(1+c)^{2}},\non[1pt]
&&Z_{1}-1=-\f{(1+2c)\lambda^{2}}{12\pi^{2}\ve c(1+c)^{2}},\non[1pt]
&&Z_{m}-1=+\f{3\lambda^{2}}{4\pi^{2}\ve c(1+c)},\label{re:counter-fermion}\eea
where all integral details are shown in Appendix \ref{app:eval-fermionself}. The main difference compared to QED is the fact that the anomalous dimension of the fermion mass is positive, i.e., $Z_{m}-1>0$. In other words, the electron mass is \textit{reduced} by spin fluctuations while it is enhanced by gauge fluctuations in QED. This is an unusual feature given by chiral gauge-field fluctuations. There exist other differences. The boson velocity appears in the renormalization factors because of the absence of the Lorentz symmetry ($c\neq v=1$). The numerator factors also differ from those of QED because there is no time-component here for chiral gauge-field fluctuations, i.e., $\Phi_{0}=0$.

The other diagrams in Fig. \ref{fig:self-energy} are for boson self-energy corrections. The calculation of Fig. \ref{fig:self-energy}-(2) is similar to that of QED while integrals of the others are in parallel with those of the $\phi^{4}$-theory. We also summarize our results only
\bea&&Z_{2}-1=-\f{\lambda^{2}}{6\pi^{2}\ve}-\f{4\Gamma_{m}}{(4\pi)^{3/2}(\ve+\et)c^{3}},\non
&&Z_{c}-1=-\f{\lambda^{2}}{6\pi^{2}\ve c^{2}},\non
&&Z_{r}-1=\f{5u}{16\pi^{2}\ve c^{3}}-\f{4\Gamma_{m}}{(4\pi)^{3/2}(\ve+\et)c^{3}},\label{re:counter-boson}\eea
where all details involved with integrals are shown in Appendix \ref{app:eval-bosonself}. The Zeeman coupling term results in negative field renormalization, which may be interpreted as screening effects for all coupling constants. The disorder scattering causes additional field renormalization, identified with additional screening effects. This additional field renormalization also decreases the boson velocity while the boson mass is unaffected by the disorder scattering since the effect on it is canceled by the $\Gamma_{m}$ term in $Z_{r}$. The boson self-interaction increases the boson mass as well known in the $\phi^{4}$-theory while it doesn't give the field renormalization in the one loop order \cite{Phi4Theory_RG}.

\subsection{Vertex corrections} \label{sec:vertex corrections}

\bf[t]\centering\ing[width=0.5\tw]{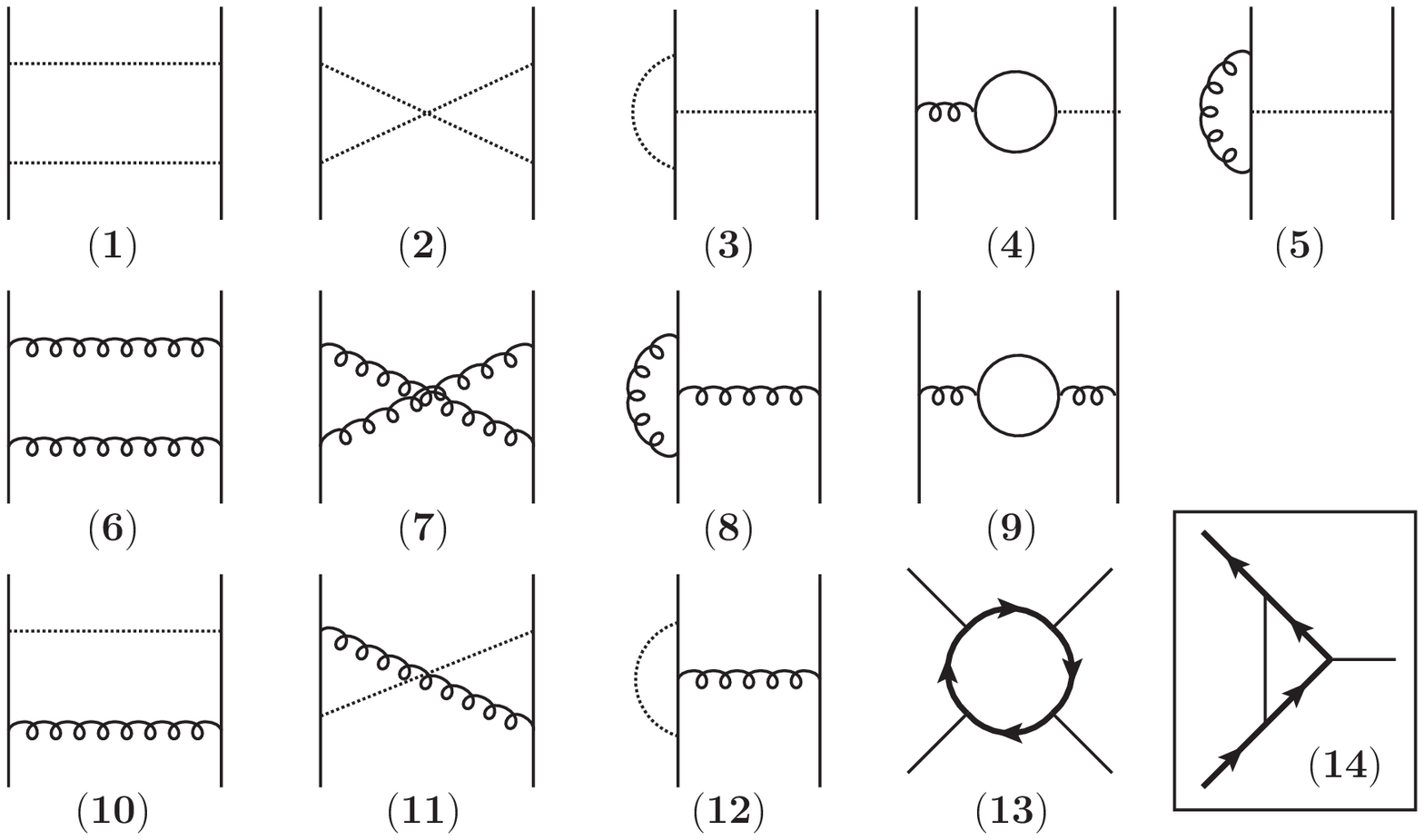}\caption{Three types of vertex corrections. Vertex corrections for the disorder scattering are in the first line, and those for the boson interaction, in the second and the third line. The last diagram in the third line should be interpreted as an amputated diagram. The diagram in the box is the correction in the one-loop order for the Zeeman coupling.}\label{fig:vertexcorrections}\ef

Vertex corrections are shown in Fig. \ref{fig:vertexcorrections}. The calculation is standard, similar to the $\phi^{4}$-theory for diagrams from Fig. \ref{fig:vertexcorrections}-(1) to Fig. \ref{fig:vertexcorrections}-(12) and QED for Fig. \ref{fig:vertexcorrections}-(13) and Fig. \ref{fig:vertexcorrections}-(14), though one should be careful about the mixing of the boson interaction and the disorder scattering, shown in Fig. \ref{fig:vertexcorrections}-(4), Fig. \ref{fig:vertexcorrections}-(5), Fig. \ref{fig:vertexcorrections}-(10), Fig. \ref{fig:vertexcorrections}-(11), and Fig. \ref{fig:vertexcorrections}-(12). Our results are summarized as
\bea&&Z_{\lambda}-1=-\f{\lambda^{2}(1+2c)}{12\pi^{2}c(1+c)^{2}\ve},\non[1pt]
&&Z_{u}-1=\f{11u}{16\pi^{2}c^{3}\ve}-\f{12\Gamma_{m}}{(4\pi)^{3/2}c^{3}(\ve+\et)},\non[1pt]
&&Z_{\Gamma_{m}}-1=-\f{8\Gamma_{m}+12u}{(4\pi)^{3/2}c^{3}(\ve+\et)}+\f{u}{4\pi^{2}c^{3}\ve}.\label{re:counter-vertex}\eea
See Appendix \ref{app:eval-vertex corrections} for more details. We point out that $Z_{\lambda}=Z_{1}$ is satisfied by the Ward identity (Sec. \ref{sec:ward identity}). The diagram Fig. \ref{fig:vertexcorrections}-(13) possibly gives renormalization for the boson interaction. However, it turns out to vanish by the Ward identity. The disorder scattering gives an antiscreening effect while the boson interaction causes a screening effect in the renormalization of the boson interaction and the disorder scattering.

\subsection{Beta functions and fixed points}\label{sec:beta functions}

\bf[t]\centering\ing[width=0.5\tw]{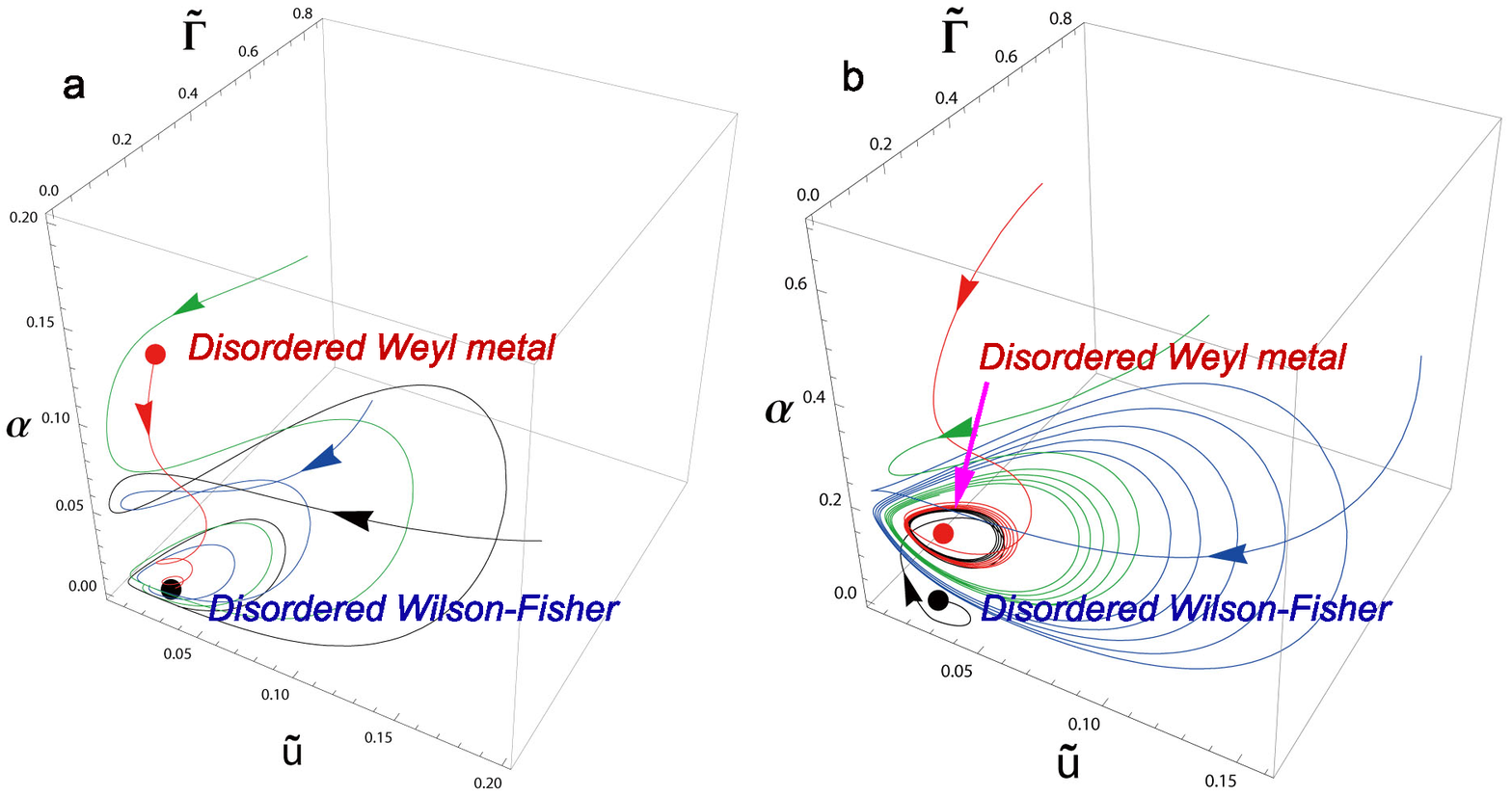}\caption{Two types of fixed points described by the beta functions in Eq. (\ref{re:beta functions}). We introduce dimensionless couplings as $\tilde{\alpha}\equiv\f{\lambda^{2}}{12\pi^{2}c}$, $\tilde{u}\equiv\f{u}{16\pi^{2}c^{3}}$, and $\tilde{\Gamma}\equiv\f{2\Gamma_{m}}{(4\pi)^{3/2}c^{3}}$. (a) When $\ve=0.01$, the boson interaction strength and the disorder strength are finite while the boson velocity and the Zeeman interaction strength vanish in the low-energy limit. This corresponds to the disordered Wilson-Fisher fixed point. (b) When $\ve=0.3$, all coupling constants are finite. This fixed point is identified as a disordered Weyl metal phase in that the fermion mass parameter gets the negative feedback from the Zeeman interaction term, and the Weyl metallic phase may arise in the presence of external magnetic fields.}\label{fig:fixedpoints}\ef

Inserting all renormalization factors of Eqs. (\ref{re:counter-fermion}), (\ref{re:counter-boson}), and (\ref{re:counter-vertex}) into the formal expressions for the beta functions Eq. (\ref{eq:beta functions}), we obtain
\bea\beta_{\lambda}&=&\lambda\bigg[-\f{\ve}{2}+\f{(c^{3}+2c^{2}+2c-4)\lambda^{2}}{12\pi^{2}c(1+c)^{2}}+\f{2\Gamma_{m}}{(4\pi)^{3/2}c^{3}}\bigg],\non
\beta_{u}&=&u\bigg[-\ve+\f{(4c^{3}+8c^{2}+10c-24)\lambda^{2}}{12\pi^{2}c(1+c)^{2}}\non
&&~~~~~~~~~~~~~~~~~~~~~~~~~~~~+\f{11u}{16\pi^{2}c^{3}}-\f{4\Gamma_{m}}{(4\pi)^{3/2}c^{3}}\bigg],\non
\beta_{\Gamma_{m}}&=&\Gamma_{m}\bigg[-1-\ve+\f{(4c^{3}+8c^{2}+12c-32)\lambda^{2}}{12\pi^{2}c(1+c)^{2}}\non
&&~~~~~~~~~~~~~~~~~~~~~~~~~~~~~~~~~~~+\f{(24\sqrt{\pi}+4)u}{16\pi^{2}c^{3}}\bigg],\non
\beta_{m}&=&m\bigg[-1+\f{(10+11c)\lambda^{2}}{12\pi^{2}c(1+c)^{2}}\bigg],\label{re:beta functions}\\
\beta_{c}&=&c\bigg[\f{(c^{4}+2c^{3}+2c^{2}-10c-1)\lambda^{2}}{12\pi^{2}c^{2}(1+c)^{2}}+\f{2\Gamma_{m}}{(4\pi)^{3/2}c^{3}}\bigg],\non
\beta_{r}&=&r\bigg[-2+\f{(c^{3}+2c^{2}+3c-8)\lambda^{2}}{6\pi^{2}c(1+c)^{2}}+\f{5u}{16\pi^{2}c^{3}}\bigg],\nn\eea
which describe the evolution of all coupling parameters as a function of an energy scale, where $\ve_{\tau} = 1$.

The limit of $\lambda \rightarrow 0$ reproduces the $\beta$-functions of the O(3) vector model with a random mass term as expected \cite{Phi4Theory_RG}. On the other hand, the existence of the $\lambda$ vertex involved with effective interactions between magnetic clusters and itinerant electrons gives rise to serious modifications on the fixed-point structure. One may point out that the $\lambda$ vertex is essentially the same as that of QED, regarded to be a U(1) gauge coupling constant. Indeed, we confirm the Ward identity, given by Sec.  \ref{sec:ward identity} in spite of the presence of the chiral matrix. However, there exists an essential different aspect between our chiral-gauge vertex and the U(1) gauge vertex of QED. The sign of the mass renormalization given by the one-loop quantum correction shows negativity instead of positivity. We recall the mass renormalization in QED, given by Ref. \cite{Phi4Theory_RG}. This ``negative" quantum correction in the chiral-gauge interaction vertex opens the possibility for the emergence of a Weyl metal phase, driven by doped magnetic impurities.

In order to verify this possibility, we solve these renormalization group equations and find two types of fixed points. They are allowed to exist until $\ve < 0.310$, which we cannot find any fixed points beyond. In the region of $0 \leq \ve < 0.0832$, the Zeeman coupling constant vanishes while both the self-interaction and disorder parameters remain finite, which is nothing but the disorder fixed point of the O(3) vector model with a random mass term. In the region of $0.0832 < \ve < 0.310$, all coupling constants are finite, identified with an interacting fixed point between itinerant electrons and doped magnetic impurities. See Fig. \ref{fig:fixedpoints}. The reason why this interacting fixed point is allowed only within this $\ve$ region is that for non-vanishing $\lambda_{*}$, $\ve$ should be large enough to overcome the screening of the disorder scattering in $\beta_{\lambda}$, i.e., $\f{\ve}{2}>\f{2\Gamma_{m*}/c_{*}^{3}}{(4\pi)^{3/2}}$ while the upper limit of $0.310$ comes from the stability condition of the fixed point. This means only in quasi-two dimensional systems does the Zeeman coupling can play a central role in low energy physical phenomena.

This nontrivial fixed point is given by
\bea&&\lambda_{*}=6.90+2.45\ln\ve,~u_{*}=4.05+0.773\ve,\label{re:rgsol1}\\
&&\Gamma_{m*}=1.03+2.45\ve,~c_{*}=1.37-0.0216\ve^{-1.5},\nn\eea
where the critical exponents near this fixed point are
\bea&&z=1.00+0.394\ve,~\eta_{\psi}=-0.00427+0.346\ve,\label{re:rgsol2}\\
&&\eta_{\Phi}=0.5\ve,~\nu_{m}^{-1}=1.08-1.99\ve,~\nu_{r}^{-1}=1.97-0.411\ve.\nn\eea
The scaling laws of both fermions and bosons are corrected. Especially, the scaling of the fermion mass ($\nu_{m}^{-1}$) is fairly nontrivial. It decreases proportionally to $\ve$ due to the screening effect of the Zeeman coupling. For example, $\nu_{m}^{-1}=1$ in the free theory becomes $\nu_{m}^{-1}=0.463$ at $\ve=0.310$. As a result, the increasing rate of the fermion mass becomes smaller. This allows an external magnetic field to overcome the mass gap.

\section{Emergence of Weyl metals from magnetically doped spin-orbit coupled semiconductors} \label{sec:emergence-Weyl}

\subsection{Applying external magnetic fields} \label{sec:field-emergence-Weyl}

\bf[t]\centering\ing[width=0.5\tw]{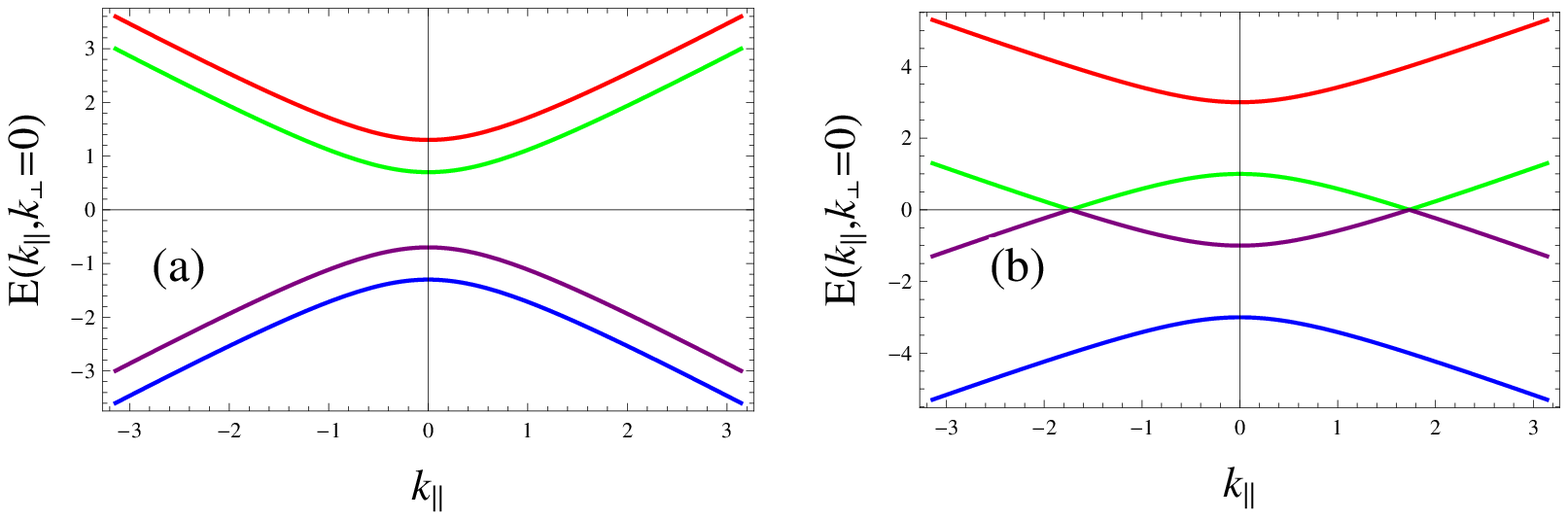}\caption{Band structures of spin-orbit coupled semiconductors with Zeeman splitting. (a) Bands are gapped when the magnetic field is smaller than the band gap while (b) a pair of Weyl points appear at the Fermi level when the magnetic field is larger than the band gap.}\label{fig:band splitting}\ef

We introduce an external magnetic field coupled to the fermions in the following way \cite{DMTS_KKM_15}
\be S_{h}=\int d^{d_{\tau}}\bm{\tau}\int d^{d}\r\bar{\psi}\bm{h}\cdot\bm{\gamma}\gamma_{5}\psi.\label{eq:ext_zeeman}\ee
This magnetic field makes two bands of two-fold degeneracy spilt into four bands as (Fig.\ref{fig:band splitting})
\be E(\k)=\pm\sqrt{\k_{\perp}^{2}+\Big(\al\bm{h}\ar\pm\sqrt{m^{2}+\k_{\parallel}^{2}}~\Big)^{2}},\label{re:band splitting}\ee
where $\k_{\parallel}=\bm{h}(\k\cdot\bm{h})/|\bm{h}|^{2}$ and $\k_{\perp}=\k-\k_{\parallel}$. When $\al\bm{h}\ar=h>m$, the mass gap is closed and a Weyl semimetal appears.

An idea is that although the mass gap is too large to be overcome by external magnetic fields, taking into account renormalization effects by doped magnetic impurities allows the gap closing as a function of the applied magnetic field and the temperature. The renormalized action of Eq. (\ref{eq:ext_zeeman}) is
\be S_{h}=\int d^{d_{\tau}}\bm{\tau}\int d^{d}\r Z_{h}\bar{\psi}_{r}\bm{h}_{r}\cdot\bm{\gamma}\gamma_{5}\psi_{r},\ee
where the renormalized magnetic field is related with the bare magnetic field as $\bm{h}_{r}=\mu^{-1}(Z_{1}/Z_{h})\bm{h}$. Then, the beta function for $h_{r}$ is given by
\be\beta_{h}=h\bigg[-1+\f{\partial\ln(Z_{h}/Z_{1})}{\partial\ln b}\bigg]=-h.\ee
We note that there is no quantum correction for $h_{r}$ in the one-loop level, consistent with the Ward identity.

Solving renormalization group equations of $\beta_{h}=-h(\mu),~\beta_{m}=-(1/\nu_{m})m(\mu),~\beta_{T}=-zT(\mu)$, we find
\be h(\mu)=h\mu^{-1},~~m(\mu)=m\mu^{-1/\nu_{m}},~~T(\mu)=T\mu^{-z},\label{re:scaling-parameters}\ee
which shows how these parameters are renormalized by the presence of doped magnetic impurities as a function of an energy scale $\mu$.

A Weyl semimetal appears when $h(\mu)>m(\mu)$, which means $h/m>\mu^{1-1/\nu_{m}}$. Replacing the scaling parameter with temperature and fixing the scale of $T(\mu)=T_{0}$, we find the gap closing condition
\be\bigg(\f{h}{m}\bigg)_{c}=\bigg(\f{T}{T_{0}}\bigg)^{\gamma},\ee
where $\gamma=\f{\nu_{m}-1}{z\nu_{m}}$. Physical units are brought back as $h=g_{s}\mu_{B}H$, where $g_{s}$, $\mu_{B}$, and $H$ are the Lande g-factor of the electron's spin, the Bohr magneton, and the external magnetic field, respectively. As a result, we obtain the critical field strength given by
\be H_{c}=(m/g_{s}\mu_{B})\bigg(\f{T}{T_{0}}\bigg)^{\gamma},\label{re:field_wm}\ee
where $\gamma=0.0828\sim0.479$ for $\ve=0.0832\sim0.310$.

\bf[t]\centering\ing[width=0.4\tw]{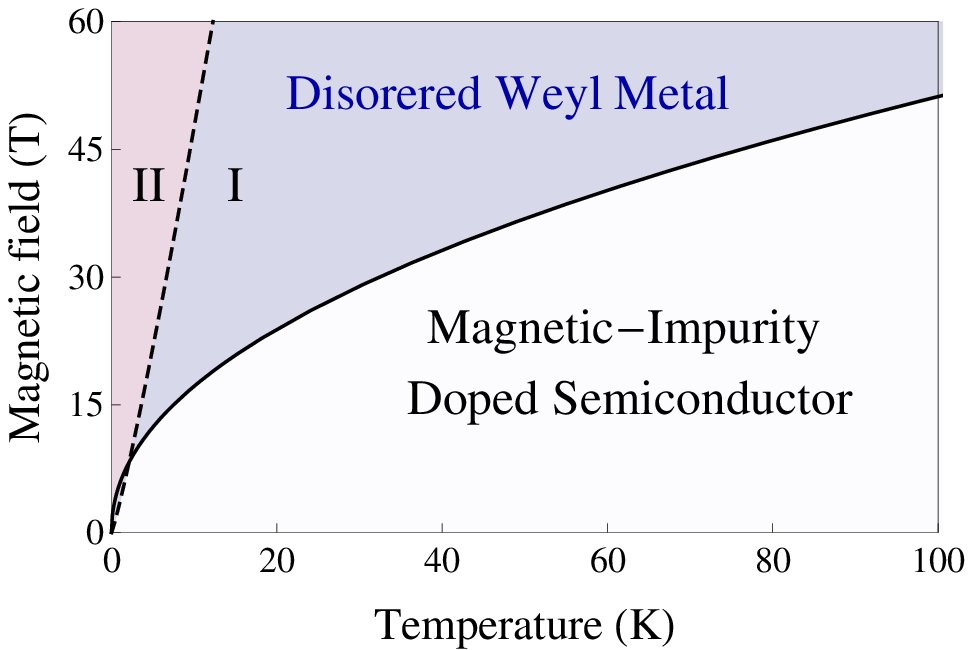}\caption{A phase diagram of dilute magnetic semiconductors in the plane of temperature and magnetic field. The solid line describes a topological phase transition from dilute magnetic spin-orbit coupled Dirac metals to non-Fermi liquid type disordered Weyl metals (I), given by Eq. (\ref{re:field_wm}). Here, the non-Fermi liquid type disordered Weyl metal phase means that Weyl electrons are incoherent due to correlations with strong antiferromagnetic spin fluctuations. The dashed line represents a crossover line from non-Fermi liquid type disordered Weyl metals (I) to Fermi liquid like conventional Weyl metals (II), where quasiparticles of Weyl electrons appear due to the suppression of antiferromagnetic fluctuations by external magnetic fields. For this crossover line, see Sec. \ref{sec:model validity}. We utilized $z=1.13$, $\nu_{m}=2.15$, $T_{0}=300~[\tx{K}]$, $m=100~[\tx{meV}]$, and $g_{s}=20$ for parameters.}\label{fig:phasediagram}\ef

Figure \ref{fig:phasediagram} shows a phase diagram based on Eq. (\ref{re:field_wm}). The external magnetic field can turn spin-orbit coupled semiconductors into Weyl semimetals by closing the band gap. The critical strength of the field is huge in the high temperature regime, for example, $H_{c}(T=300~[\tx{K}])= 86.2 ~[\tx{T}]$. However, $H_{c}(T)$ becomes much smaller at low temperatures because the mass gap gets screened by spin fluctuations while the magnetic field is unaffected. Using $T_{0}=300~[K]$ as an UV energy scale and $m= 100 ~[meV]$, $g_{s} = 20$ for typical values at that scale, we find
\be H_{c} = 9.50\sim58.9 ~ [T]~(\tx{at }T=3~[K]),\label{re:field_wm_number}\ee
for $\ve=0.0832\sim0.310$.

\subsection{Negative longitudinal magneto-resistivity} \label{sec:NLMR}

\bf[t]\centering\ing[width=0.4\tw]{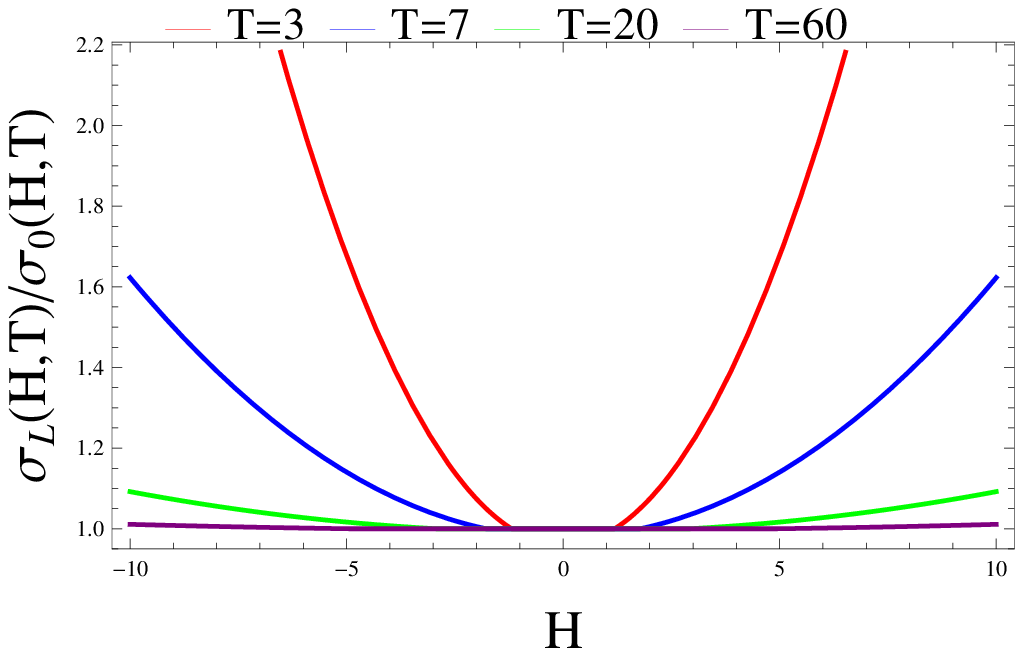}\caption{Longitudinal magnetoconductivities in Weyl semimetals. The conductivities are normalized with the Drude conductivity. For parameters, $a=0.2$ and $b=0.1$ are used.}\label{fig:conductivity}\ef

The existence of a topological phase transition from either a topological or band insulating state to a Weyl metal phase has been confirmed by our renormalization group analysis. Although this demonstration itself touches a novel aspect in the study of dilute magnetic semiconductors, it is necessary to verify the present scenario more quantitatively. Here, we focus on the longitudinal magneto-resistivity, which acquires an essential modification in the Weyl metal state, given by
\be\sigma_{L}(H,T)=\sigma_{0}(T)(1+C_{W}H^{2}),\label{eq:lmc}\ee
where $\sigma_{0}(T)$ is the Drude conductivity with a weak anti-localization quantum correction and $C_{W}$ is a positive coefficient with the applied magnetic field along the direction of the electric field \cite{WM_Boltzmann_KSK}. This modification has been proposed to originate from the chiral anomaly \cite{Chiral_Anomaly_I,Chiral_Anomaly_II,Chiral_Anomaly_III,Chiral_Anomaly_IV}, where $C_{W}H^{2}\sim h^{2}-m^{2}$ corresponds to the square of the momentum-space distance between a pair of Weyl points \cite{DMTS_KKM_15,WM_Boltzmann_KSK,Anomaly_Chiral_Current} given by $\k_{\parallel}=\sqrt{h^{2}-m^{2}}$ from Eq. (\ref{re:band splitting}).

Introducing renormalization effects from renormalization group equations into this expression, given by Eq. (\ref{re:scaling-parameters}), we find
\be C_{W}(H,T)=\f{a}{T^{2/z}}-\f{b}{H^{2}T^{2/z\nu_{m}}},\label{re:cw}\ee
where $a\propto g_{s}^{2}$ and $b\propto m^{2}$ are regarded to be phenomenological fitting parameters \cite{Experiments_Doping_MI}. Note that $C_{W}$ depends not only temperature but also external magnetic fields. This is not the case in usual Weyl semimetals, where $C_{W}$ is just a coefficient of the negative MR \cite{WMD_KKM_16}. This is a characteristic feature of Weyl semimetals arising from magnetically doped spin-orbit coupled semiconductors \cite{Experiments_Doping_MI}.

Figure \ref{fig:conductivity} shows longitudinal conductivities normalized with the Drude conductivity, given by
\be\sigma_{L}(H,T)/\sigma_{0}(H,T)=1+\bigg(\f{a}{T^{1.78}}-\f{b}{H^{2}T^{0.838}}\bigg)H^{2},\label{re:LMC}\ee
where we used $z=1.13$ and $\nu_{m}=2.15$ for a numerical estimate. Positive magneto-conductivities start at $H_{c}=1.2,~1.8,~2.9,~4.9$ for $T=3,~7,~20,~60$, respectively. Moreover, the positive magneto-conductivity is significantly enhanced as temperature is lowered.

\section{Discussion} \label{sec:discussion}

\subsection{Discussion on our model construction} \label{sec:model validity}

\subsubsection{Role of an effective Zeeman interaction between external magnetic fields and antiferromagnetic spin fluctuations}

One may criticize that our effective field theory does not take into account an effective Zeeman interaction between external magnetic fields and antiferromagnetic spin fluctuations. He/she may claim that the absence of the effective Zeeman coupling gives rise to a severe conceptual flaw of the present study. The main conclusion is that magnetic fluctuations of dopant atoms near an antiferromagnetic transition effectively (in the sense of effective field theory) decrease the bulk insulating gap, which lowers the energy scale of perturbations required to reach a Weyl semi-metal phase. However, the magnetic field which is producing the Weyl semi-metal phase should itself suppress antiferromagnetic fluctuations and may therefore counteract the effect of the magnetic field claimed by the present study.

Taking into account this effective Zeeman interaction, we have a modified potential energy for a ferromagnetic cluster, $V(\P)=-g_{\Phi}\mu_{\Phi}\bm{H}\cdot\P+r\P^{2}+\f{u}{8}(\P^{2})^{2}$, where $\mu_{\Phi}$ and $g_{\Phi}$ are the magnetic moment and the Lande-g factor of the ferromagnetic cluster, respectively. The magnetic field lowers the potential energy by making $\P$ point in its direction, suppressing fluctuations of $\P$ in the ground state. At finite temperatures, effects of entropy making $\P$ disordered compete with the potential energy. It is natural to consider a crossover temperature, given by $T_{mag}=C_{1}g_{\Phi}\mu_{\Phi}|\bm{H}|^{1/z}$, where $C_{1}$ is a positive constant and $z$ is the dynamical critical exponent. The breakdown of the $H/T$ scaling is an essential property from our renormalization group analysis. Above this crossover temperature, antiferromagnetic fluctuations are still strong, responsible for the negative feedback effect on the carrier gap. In other words, there exists an intermediate temperature regime $T_{mag}<T<T_{c}$ for a given $\bm{H}$, where $T_{c}$ is the temperature scale for a topological phase transition from dilute magnetic spin-orbit coupled Dirac metals to non-Fermi liquid type disordered Weyl metals, given by Eq. (\ref{re:field_wm}). Below the crossover temperature $T_{mag}$, such antiferromagnetic fluctuations are suppressed and ferromagnetic components are expected to appear. As a result, a conventional Weyl metal phase would be realized, where quasiparticles of Weyl electrons exist. This consideration introduces an additional phase boundary of $H_{mag}=\Big(\f{T}{C_{1}g_{\Phi}\mu_{\Phi}}\Big)^{z}$ in the phase diagram of Fig. \ref{fig:phasediagram}. Considering $H_{mag} = H_{c}$, where $H_{c}$ is given by Eq. (\ref{re:field_wm}), we obtain a threshold value for the Lande-g factor, given by $g_{\Phi}^{mag} = \f{T_{0}}{C_{1}\mu_{\Phi}} \Big(\f{g_{s}\mu_{B}}{m}\Big)^{1/z}$ with $T \approx T_{0}$. When the Lande-g factor is larger than this threshold value, the negative feedback effect disappears by the magnetic field effect to suppress spin fluctuations. Using the same parameter for Eq. (\ref{re:field_wm}), we find that this happens when $g_{\Phi}\geq\f{1.95}{C_{1}\mu_{\Phi}}$.

\subsubsection{Role of the existence of a small Fermi surface}

One may ask the origin of effective interactions between doped magnetic ions. We would like to emphasize that our effective UV lattice model is given by an O(3) Heisenberg model with random exchange interactions as our starting point. Frankly speaking, the physically meaningful mathematical description of this effective spin interaction term is not completely clear at all. This is the reason why we explain our effective UV lattice model in a very careful sentence, ``Although this aspect does not mean that the Ruderman-Kittel-Kasuya-Yosida (RKKY) interaction would be the mechanism of effective interactions between doped magnetic moments, where the dynamics of such electrons with a small Fermi surface would be essential to determine the nature of their effective interactions, we resort to this physical picture as our reference." However, the question on the role of the small Fermi surface would be quite relevant for possible consistency between the UV and infrared (IR) physics for this problem although we focused on the semi-metallic regime in this study.

Now, we take a chemical potential slightly above the mass gap in a semi-metallic regime ($\mu\sim m$). Then, we find that the result is not much changed in the present level of approximation if we do not consider possible damping effects on the boson dynamics. We refer all details to Appendix \ref{app:fermi surface}. On the other hand, we point out that that the finite density of fermions gives rise to a Landau damping term for bosons, given by $\mu^{2}\f{|q_{0}|}{|\q|}$. This Landau damping term becomes more singular than the quadratic term $q_{0}^{2}$ in the low frequency regime. We also mention that this Landau damping term can be modified when disorder scattering is introduced into fermions directly, resulting in a diffusive fixed point for the dynamics of itinerant electrons of the bulk. But, the ``pseudogap" effect allows an effective ``ballistic regime" in the most temperature regime, where the small Fermi surface may result in the diffusive dynamics of electrons at a relatively low temperature regime. In this respect the above Landau damping term can be taken into account. For related discussions, see section \ref{sec:random mass}. Then, our renormalization group analysis will lose its validity in that regime. This consideration introduces another crossover scale into the result. We estimate the crossover scale based on $q_{0}^{2}\sim T^{2}$ and $\mu^{2}|q_{0}|/|\q|\sim\mu^{2}$ since the dynamical critical exponent $z$ is almost one, so $|q_{0}|/|\q|\sim T^{z}/T\sim1$. When $T\gg T_{\mu}=C_{2}\mu$ with a positive constant $C_{2}$, our renormalization group analysis is applicable.

\subsubsection{On the effective field theory for antiferromagnetic fluctuations}

Although the O(3) vector model (IR effective field theory) has been derived from the O(3) Heisenberg model (UV effective Hamiltonian) in the O(3) symmetric paramagnetic vacuum, one may consider an alternative description for the spin dynamics of antiferromagnetic fluctuations. For example, an O(3) nonlinear $\sigma$ model can be suggested, which would be derived from the Heisenberg model in an O(3) symmetry broken ground state.

An O(3) vector model is an effective field theory for a phase transition characterized with O(3) symmetry breaking, which starts from an O(3) symmetric state. On the other hand, a nonlinear $\sigma$ model is an alternative description for the same phase transition, but it starts from a symmetry broken state. We recall the Haldane mapping from the O(3) Heisenberg model to an effective O(3) nonlinear $\sigma$ model with a theta term \cite{Haldane_Mapping}. In other words, field contents of the nonlinear $\sigma$ model describe elementary excitations from the symmetry broken ground state. It contains phase or (angular or transverse) fluctuations of the order parameter but neglects amplitude (radial or longitudinal) fluctuations, which become massive so less important in the ordered phase. Here,``an alternative description for the same phase transition" means that performing the renormalization group analysis for the nonlinear $\sigma$ model, we reach the same critical point as described by the Landau-Ginzburg field theory, for example, the Wilson-Fisher fixed point. An explicit demonstration can be found in Ref. \cite{Peskin}, where O(2) symmetry is considered.

In this study we start from a paramagnetic state at high temperatures and approach the critical point, where order-parameter fluctuations proliferate. We investigated nontrivial effects of such fluctuations on strong spin-orbit coupled electrons. In this respect the O(3) vector model seems more appropriate than the nonlinear $\sigma$ model for our problem. Meanwhile, we might start from a nonlinear $\sigma$ model because these two models are expected to have the same universal physics. Unfortunately, the appearance of the gradient coupling interaction between itinerant electrons and Goldstone boson excitations, referred to as Adler's principle \cite{Adler_Principle}, does not affect the dynamics of itinerant electrons at least in the one-loop order renormalization group analysis. This is the reason why Landau's Fermi-liquid state appears even in the symmetry broken phase although there exist gapless Goldstone boson excitations. Of course, one may describe possible strong coupling physics between fermions and Goldstone bosons at a critical point beyond the one-loop order based on the nonlinear $\sigma$ model description, but we do not know any explicit calculations.

\subsubsection{Connection between the disorder strength and dopant concentration}

One may ask how the disorder strength, given by the variance of the random mass in this study, is related with dopant concentration. It is natural to assume that the effective exchange interaction between doped magnetic ions in average is proportional to the average distance between magnetic ions, given by $\sim n_{imp}^{-1/3}$ in three dimensions, where $n_{imp}$ is the concentration of magnetic dopants. An essential point is how the variance of the exchange interaction can be related with the variance of the impurity position or the distance between magnetic ions. Here, we assume that they are proportional to each other. Then, the final question is how the variance of the distance between magnetic ions is related with the impurity concentration. Applying the central limit theorem to this situation, we conjecture $\sqrt{\langle\delta r_{ij}^{2}\rangle} \sim n_{imp}^{-1/2}$ in the thermodynamic limit. If we assume $\sqrt{\langle\delta J_{ij}^{2}\rangle} \sim \sqrt{\langle\delta r_{ij}^{2}\rangle}$ as discussed above, we suggest $\sqrt{\langle\delta J_{ij}^{2}\rangle} \sim n_{imp}^{-1/2}$. However, we cannot give any rigorous arguments for this monotonic behavior with respect to the impurity density.

\subsection{Origin of the negative feedback effect} \label{sec:negative feedback}

To clarify the physical picture, we demonstrate how chiral gauge fields given by spin fluctuations cause the negative feedback effect of the mass gap.
%
%
First, we consider the absence of spin fluctuations. When a mass term is zero ($m=0$), Dirac fermions can be described by their chiral states, denoted by $\left|+,\uparrow,\k\right\rangle$ and $\left|+,\downarrow,\k\right\rangle$ for $``+"$-chiral fermions and $\left|-,\uparrow,\k\right\rangle$ and $\left|-,\downarrow,\k\right\rangle$ for $``-"$- chiral fermions whose eigenvalues are $|\k|$, $-|\k|$, $-|\k|$, and $|\k|$, respectively. Here, up- and down-spin states denote diagonal bases of spin-orbit coupling for each chiral block, where the spin direction varies with $\k$. Now, we turn on a mass term. The mass term mixes two chiral states. For example, it describes scattering from $\left|+,\uparrow,\k\right\rangle$ into $\left|-,\uparrow,\k\right\rangle$. Actually, the overlap of both wave-functions is given by $\left\langle+,\sigma,\k\right|mI_{2}\left|-,\sigma,\k\right\rangle=m$, where $I_{2}$ is a two-by-two unit matrix applied to the chiral block. This overlap integral determines the size of the excitation gap, given by the energy spectrum as $E_{\k}$, $-E_{\k}$, $-E_{\k}$, and $E_{\k}$ with $E_{\k}=\sqrt{|\k|^{2}+m^{2}}$.

Next, we take into account magnetic fluctuations. We recall that such spin fluctuations are described by chiral gauge fields. When chiral gauge fields are present, they reduce the wave-function overlap. Here, we consider the case of uniform fields for simplicity. Now, chiral states are modified as $\left|+,\sigma,\k+\P\right\rangle$ and $\left|-,\sigma,\k-\P\right\rangle$, where $\P$ represents the uniform chiral gauge field. Note the sign difference in front of $\P$. Spin states of two chiral states are rotated in opposite directions. This difference makes their overlap less than unity, resulting in $\left\langle+,\sigma,\k+\P\right|mI_{2}\left|-,\sigma,\k-\P\right\rangle<m$. Expanding the overlap with $\P$, we find reduction of the effective mass gap as follows
\be m_{eff}=m\left[1-\f{(\partial_{\theta_{\k}}\bm{\hat{k}}\cdot\bm{\Phi})^{2}}{|\k|^{2}} - \f{(\partial_{\phi_{\k}}\bm{\hat{k}}\cdot\bm{\Phi})^{2}}{(|\k|\sin{\theta_{\k}})^{2}}\right]^{1/2} \ee
with $\bm{\hat{k}}=\k/|\k|$, where $\theta_{\k}$ and $\phi_{\k}$ are the polar angle and the azimuthal angle of $\bm{\hat{k}}$, respectively. This is in contrast to the case of gauge fields. A vector field $\A$ doesn't change the wave-function overlap, which rotates two chiral states equivalently, resulting in $\left\langle+,\sigma,\k-\A\right|mI_{2}\left|-,\sigma,\k-\A\right\rangle=m$. The mass gap is not affected by the gauge field in this level of approximation.

We would like to point out that the chiral field was taken into account in a perturbative way but the mass term was treated exactly. If we take both terms on equal footing, we will have band splitting similar with Eq. (\ref{re:band splitting}) instead of the reduced gap. We recall that the chiral gauge field of the short distance scale has been integrated out in the renormalization group analysis, where the chiral gauge field of that scale is smaller than the mass gap. In this respect taking into account the chiral gauge field in a perturbative way is actually what we performed in the renormalization group approach.

If we should take into account fluctuations of chiral gauge fields, the task is not straightforward anymore. We need to count wave-function renormalization factors and vertex corrections. All of these renormalization factors were taken into account systematically in the renormalization group approach, where the gap-reduction is manifest in the reduction of the scaling exponent of the carrier gap.

\subsection{Role of the random-mass disorder} \label{sec:random mass}

One cautious person may point out that the random mass of magnetic modes does not seem to be crucial for the present phenomenon. Actually, introduction of the randomness reflects the real physical situation. The softening of the carrier gap is the direct result of the effective Zeeman coupling between fluctuating local magnetic moments and Dirac electrons. To clarify this physics, we set $\Gamma_{m}=0$ in Eq. (\ref{re:beta functions}). Then, we obtain
\bea\beta_{c}&=&c\bigg[\f{(c^{4}+2c^{3}+2c^{2}-10c-1)\lambda^{2}}{12\pi^{2}c^{2}(1+c)^{2}}\bigg],\non
\beta_{\lambda}&=&\lambda\bigg[-\f{\ve}{2}+\f{(c^{3}+2c^{2}+2c-4)\lambda^{2}}{12\pi^{2}c(1+c)^{2}}\bigg],\non
\beta_{u}&=&u\bigg[-\ve+\f{(4c^{3}+8c^{2}+10c-24)\lambda^{2}}{12\pi^{2}c(1+c)^{2}}+\f{11u}{16\pi^{2}c^{3}}\bigg],\non
\beta_{m}&=&m\bigg[-1+\f{(10+11c)\lambda^{2}}{12\pi^{2}c(1+c)^{2}}\bigg].\label{re:beta functions-clean}\eea
When $\ve>0$, we have a fixed point given by
\be c_{*}=1.49,~\lambda_{*}=9.03\sqrt{\ve},~u_{*}=0,\label{re:clean}\ee
where critical exponents are
\be z=1+0.377\epsilon,~~~\nu_{m}^{-1}=1-1.97\epsilon . \ee
These are almost same with those of Eq. (\ref{re:rgsol2}). In this respect all physical phenomena in Sec. \ref{sec:emergence-Weyl} remains to be valid in the clean case. For example, the critical field strength for the gap closing is given by $H_{c}\sim T^{\gamma}$ with $\gamma=0\sim0.548$ for $\ve=0\sim0.3$.

In the absence of the random-mass term, the negative feedback effect exists for all values of $\ve$ as long as $\ve>0$. In the presence of the random-mass term, such an effect exists only in $0.0832<\ve<0.310$. Thus, the role of the random-mass term affects the negative feedback effect indirectly by reducing the range of $\ve$. This means that the negative feedback effect appears more easily in homogeneous systems. However, there always exist certain amount of disorders in actual systems of magnetically doped semiconductors. In this case the clean fixed point in Eq. (\ref{re:clean}) becomes unstable (see Table. \ref{tab:fixedpoints}), and the disordered fixed point in Eq. (\ref{re:rgsol1}) will appear to govern real physical phenomena.

\subsection{Role of potential scattering in the emergent disordered Weyl metal phase} \label{sec:fate of weyl metal}

One cautious person may criticize that the role of potential scattering has been neglected in our emergent disordered Weyl metal phase. If the translational symmetry is broken, scattering between Weyl nodes could potentially mix the chirality of Weyl fermions and destroy the topological protection. Then, one can ask whether the disordered Weyl metal phase can survive in the presence of impurity scattering or not. In this section, we show that there exists a weak-scattering regime, where the disordered Weyl metal phase remains stable.

We recall Eq. (\ref{re:band splitting}), which shows a pair of Weyl points at $\k=+\c$ and $\k=-\c$, where $\c=\bm{h}\sqrt{1-(m/|\bm{h}|)^{2}}$. Now, we expand fermion fields near the Weyl points as $\psi(\tau,\r)=\sum_{\k}\psi_{+}(\tau,\r)e^{\i(\k+\c)\cdot\r}+\sum_{\k}\psi_{-}(\tau,\r)e^{\i(\k-\c)\cdot\r}$. Then, we obtain an effective Weyl-fermion action, given by
\be S=\int^{\beta}_{0}d\tau\int d^{d}\r\bar{\Psi}(\partial_{\tau}\gamma_{0}-\i\partial_{\r}\cdot\bg+\c\cdot\bg\gamma_{5})\Psi,\ee
where $\Psi=(\psi_{+},\psi_{-})^{T}$. Next, we consider nonmagnetic impurity scattering on Weyl fermions, given by
\bea S_{dis}&=&-\int^{\beta}_{0}d\tau\int^{\beta}_{0}d\tau'\int d^{d}\r\f{\Gamma_{1}}{2}\bar{\Psi}_{\tau}\Psi_{\tau}\bar{\Psi}_{\tau'}\Psi_{\tau'}\non
&&-\int^{\beta}_{0}d\tau\int^{\beta}_{0}d\tau'\int d^{d}\r\f{\Gamma_{2}}{2}\bar{\Psi}_{\tau}\gamma_{0}\Psi_{\tau}\bar{\Psi}_{\tau'}\gamma_{0}\Psi_{\tau'},\non\eea
where $\Gamma_{1}$ ($\Gamma_{2}$) is the disorder strength for inter-valley (intra-valley) scattering.

Two of us have performed the renormalization group analysis for this effective field theory, where renormalization group equations for the distance between the pair of Weyl points $\bm{c}$ and both disorder scattering parameters $\Gamma_{1}$ and $\Gamma_{2}$ are derived to describe their evolutions with respect to temperature \cite{WMD_KKM_16}. Here, self-energy corrections are introduced in the two-loop order while disorder vertex corrections are taken into account up to the one-loop order. The intra-valley scattering strength remains irrelevant, regarded to be the pseudogap effect of the Weyl band structure. On the other hand, the inter-valley scattering strength turns out to flow into a weak disorder fixed point, the existence of which results from the self-energy correction of the two-loop order. An interesting and unexpected result is that the effect of inter-valley scattering results in positive renormalization for the distance between the Weyl pair, i.e., $|\c|$, as temperature lowered. This gives rise to more rapid enhancement of the Weyl-pair distance than that of the clean case.

Combining the scaling theory of Eq. (\ref{re:field_wm}) with the result of Ref. \cite{WMD_KKM_16}, we obtain
\be|\c|\sim (T/T_{c})^{a}\sqrt{1-(H_{c}/H)^{2}(T/T_{c})^{2b}},\ee
where $b=\f{\nu_{m}-1}{z\nu_{m}}=0.166\sim0.957$ for $\ve=0.0832\sim0.310$. $a$ is a function of the fixed point value of $\Gamma_{1}$, where we obtain $a=-1$ for $\Gamma_{1}=0$. If $a$ is positive, $|\c|$ will decrease in the case of $T<T_{c}\big[(H/H_{c})\sqrt{b/(a+b)}\big]^{1/b}$ and vanish eventually as $T\rightarrow0$. On the other hand, if $a$ is negative, the Weyl metal phase will persist in a low-temperature regime. The perturbative renormalization group study revealed $a=-1.6$, showing more rapid enhancement of the Weyl-pair distance than that of the clean case due to the inter-valley scattering \cite{WMD_KKM_16}. Based on this discussion, we claim that the disordered Weyl metal phase remains stable at least in the weak-scattering regime.

\subsection{Ward identity} \label{sec:ward identity}

In our renormalization group analysis we observed that the Ward identity of $Z_{1}=Z_{\lambda}$ is satisfied. This is rather unexpected in the respect that the chiral symmetry is explicitly broken by the mass term in the classical level. However, the result turns out to be consistent with the symmetry analysis based on the Schwinger-Dyson equation \cite{Peskin}. Here, we briefly sketch the derivation. See Appendix \ref{app:ward identity} for details.

We start from the fermion Green's function, given by
\be\left\langle\psi(x_{1})\bar{\psi}(x_{2})\right\rangle=\f{1}{\Z}\int\D(\bar{\psi},\psi)\psi(x_{1})\bar{\psi}(x_{2})e^{-S},\label{eq:fermion green}\ee
where $\Z = \int\D(\bar{\psi},\psi) e^{-S}$ is the partition function. In this expression we focus on the fermion action of $S_{f}=\int dx\bar{\psi}(x)(\i\partial_{\mu}\gamma_{\mu}+m)\psi(x)$ with $\partial_{\mu}=(-\i\partial_{0},\partial_{\r})$ and $\gamma_{\mu}=(\gamma_{0},\bm{\gamma})$.

One may consider the chiral symmetry of Eq. (\ref{eq:fermion green}) based on the chiral transformation
\be\psi(x)\rightarrow e^{\i\alpha(x)\gamma_{5}}\psi(x)\label{eq:chiral transformation}.\ee
Inserting Eq. (\ref{eq:chiral transformation}) into Eq. (\ref{eq:fermion green}), we organize the resulting expression order by order in $\alpha(x)$. In the first order of $\alpha(x)$, we obtain
\bea k_{\mu}\Gamma_{\mu 5}(p+k,p)&=&[G^{-1}(p+k)-G^{-1}(p)]\gamma_{5}\non
&&+2m[\Gamma_{5}(p+k,p)-Z_{m}\gamma_{5}],\eea
where $\Gamma_{\mu 5}(p+k,p)$ is the one-particle irreducible vertex for $\int dp_{1}\left\langle\bar{\psi}(p_{1}+k)\gamma_{\mu}\gamma_{5}\psi(p_{1})\psi(q)\bar{\psi}(p)\right\rangle$ and $\Gamma_{5}(p+k,p)$, that for $\int dp_{1}\left\langle\bar{\psi}(p_{1}+k)\gamma_{5}\psi(p_{1})\psi(q)\bar{\psi}(p)\right\rangle$. The vertex function $\Gamma_{\mu5}$ arises from the chiral current generated by the chiral transformation while the fermion Green's function results from nontrivial commutation of fields \cite{Peskin}. There appear new terms, identified with the pseudo-scalar term $\Gamma_{5}$ and the mass term $Z_{m}$, both of which originate from the explicit symmetry breaking given by the mass term. In Appendix \ref{app:ward identity} we confirm that these two terms are cancelled to each other in the renormalization group structure. As a result, we obtain the Ward identity:
\be Z_{\mu 5}\gamma_{\mu}\gamma_{5}=\f{\partial G^{-1}(p)}{\partial p_{\mu}}\gamma_{5},\label{eq:ward identity}\ee
where $Z_{\mu5}\gamma_{\mu}\gamma_{5}\equiv\lim_{k\rightarrow0}\Gamma_{\mu 5}(p+k,p)$.

This Ward identity forces the relationship among the renormalization factors, resulting in $Z_{1}=Z_{\lambda}=Z_{h}$. Spin fluctuations reduce the fermion's velocity and renormalize the coupling with the U(1) chiral-current. However, these two effects are exactly cancelled to each other. As a result, the coupling with the U(1) chiral-current, associated with the chiral symmetry, is not affected by quantum corrections given by chiral gauge fields. In other words, the external magnetic field does not have an anomalous dimension.

\subsection{Higher order quantum corrections} \label{sec:higher order}

\bf[t]\centering\ing[width=0.4\tw]{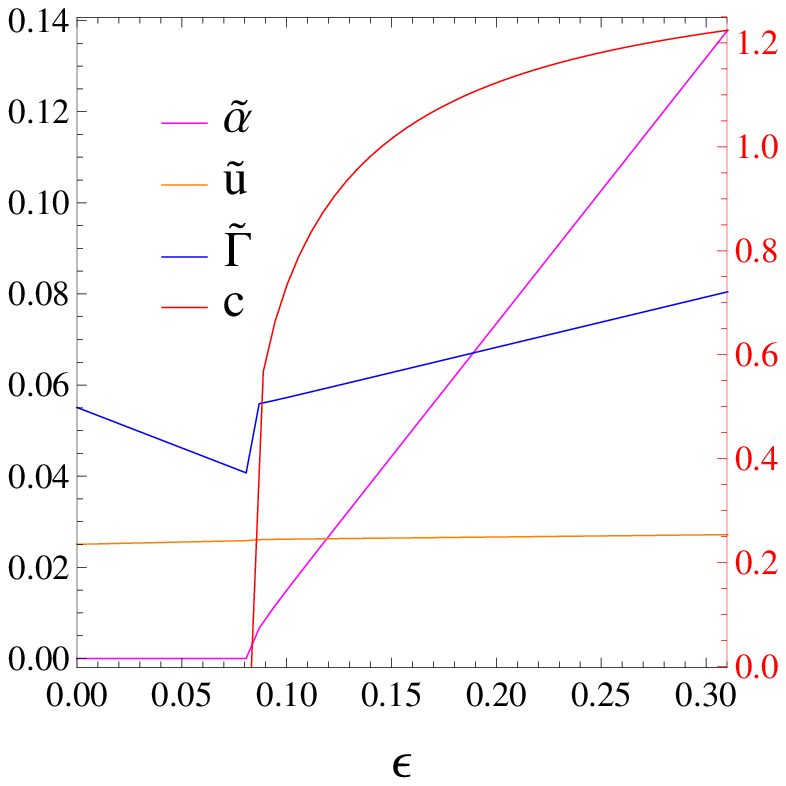}\caption{Fixed point values of the dimensionless coupling constants ($\tilde{\alpha}$, $\tilde{u}$, and $\tilde{\Gamma}$; left axis) and the boson velocity ($c$; right axis) are shown as a function of $\ve$.}\label{fig:couplings}\ef

We show how our one-loop renormalization group result can be justified, including higher-order quantum corrections. More precisely, we figure out how general Feynman diagrams depend on both coupling constants and boson velocity. We consider a general $L$-loop diagram, given by
\bea I\sim u^{V_{1}}\Gamma_{m}^{V_{2}}\lambda^{V_{3}}\int\left[\prod_{i=1}^{L}dp_{i}\right]\prod_{l=1}^{I_{b}}\left\{\f{1}{\q_{\tau,l}^{2}+c^{2}\q_{l}^{2}+r}\right\}\non\times\prod_{j=1}^{I_{f}}\left\{\f{1}{-\i\k_{\tau,j}\cdot\bg_{\tau,j}+\k_{j}\cdot\bg+m}\right\},\eea
where $k_{j}$ and $q_{l}$ are linear combinations of the internal momenta $p_{i}$ and the external momenta. $V_{1}$, $V_{2}$, and $V_{3}$ are the number of interaction vertices, given by the boson self-interaction $u$, the random-mass induced interaction $\Gamma_{m}$, and the fermion-boson interaction $\lambda$, respectively. $L$ is the total number of loops, and $I_{f}$ ($I_{b}$) is the number of fermion (boson) propagators.

To estimate the dependence on the velocity $c$, we note that there are three types of loops: loops made of boson propagators (boson-loops), loops made of fermion propagators (fermion-loops), and loops made of both propagators (boson-fermion loops). Boson-loops appear from the boson self-interaction and the disorder scattering (see Fig. \ref{fig:self-energy}-(3)) while fermion-loops and boson-fermion loops appear from the Zeeman coupling (see Fig. \ref{fig:self-energy}-(2) and Fig. \ref{fig:self-energy}-(1), respectively). We consider a case where there are $L_{b}$ boson-loops, $L_{f}$ fermion-loops, and $L_{bf}$ boson-fermion loops. The total number of loops is given by $L=L_{b}+L_{f}+L_{bf}$. For the boson-loops, we find a factor of  $c^{-3L_{b}}$ from the scaling of $q_{l}\rightarrow q_{l}/c$. For the boson-fermion loops, the same scaling analysis gives rise to $c^{-L_{bf}}H(c,v)$, where $H(c,v)$ describes mixing between the boson velocity and the fermion velocity, here, $v=1$, with $H(c_{*},1) \sim \O(1)$ (see Eq. (\ref{re:counter-fermion}), for example). For the fermion-loops, we don't have any $c$-factor because there are no boson propagators in the loops. Totally, we find the boson-velocity factor as $c^{-3L_{b}-L_{bf}}H(c,v)$ in this loop diagram.

Next, we rewrite the general loop integral $I$ in terms of  the dimensionless coupling constants, introduced before and given by
\be\tilde{\alpha}\equiv\f{\lambda^{2}}{12\pi^{2}c},~\tilde{u}\equiv\f{u}{16\pi^{2}c^{3}},~\tilde{\Gamma}\equiv\f{2\Gamma_{m}}{(4\pi)^{3/2}c^{3}}.\label{eq:dimensionless couplings}\ee
Then, we obtain \be I\sim\tilde{u}^{V_{1}}\tilde{\Gamma}^{V_{2}}\tilde{\alpha}^{V_{3}/2}c^{\rho}H(c,1),\ee where we introducd $\rho=3V_{1}+3V_{2}+V_{3}/2-(3L_{b}+L_{bf})$. Using the following identities of $L=I_{b}+I_{f}-(V_{1}+V_{2}+V_{3})+1$, $4(V_{1}+V_{2})+V_{3}=2I_{b}+E_{b}$, and $2V_{3}=2I_{f}+E_{f}$, where $E_{b}$ ($E_{f}$) is the number of the external boson (fermion) lines, we simplify $\rho$ as $\rho=2\delta+\f{1}{2}(E_{b}+E_{f}-2)+L_{f}$ with $\delta=V_{1}+V_{2}-L_{b}\geq0$. Based on this result, we find all the renormalization factors in the $L$-loop order as follows
\bea\delta_{0},\delta_{1},\delta_{m},\delta_{2},\delta_{r}&\sim&\tilde{u}^{V_{1}}\tilde{\Gamma}^{V_{2}}\tilde{\alpha}^{V_{3}/2}c^{2\delta+L_{f}},\non
\delta_{c}&\sim&\tilde{u}^{V_{1}}\tilde{\Gamma}^{V_{2}}\tilde{\alpha}^{V_{3}/2}c^{2\delta+L_{f}-2},\non
\delta_{\lambda}&\sim&\tilde{u}^{V_{1}}\tilde{\Gamma}^{V_{2}}\tilde{\alpha}^{(V_{3}-1)/2}c^{2\delta+L_{f}},\non
\delta_{u}&\sim&\tilde{u}^{V_{1}-1}\tilde{\Gamma}^{V_{2}}\tilde{\alpha}^{V_{3}/2}c^{2\delta+L_{f}-2},\non
\delta_{\Gamma_{m}}&\sim&\tilde{u}^{V_{1}}\tilde{\Gamma}^{V_{2}-1}\tilde{\alpha}^{V_{3}/2}c^{2\delta+L_{f}-2}.\label{eq:higher order}\eea
Note that all the renormalization factors are proportional to positive powers of the couplings. If the couplings are small, so are the renormalization factors.

Figure \ref{fig:couplings} shows the fixed point values of the couplings obtained in the one-loop order. The boson velocity is the order of unity ($c_{*}\leq1.2$) and all other couplings are much less than unity, i.e., $\tilde{u}_{*}\simeq0.025$, $\tilde{\Gamma}_{*}\leq0.08$, and $\tilde{\alpha}_{*}\leq0.12$, respectively. As a result, the renormalization factors are more strongly suppressed in higher orders. This demonstration supports an idea that the fixed point solution that we obtained in the one-loop order remains valid, even including higher-order quantum corrections. However, we admit that it doesn't prove the idea obviously since it is not clear whether the total sum converges or diverges. We stress that other factors such as symmetry factors and momentum integrations, which we didn't take into account explicitly in Eq. (\ref{eq:higher order}), don't give rise to any enhancement factor.

\section{Summary} \label{sec:summary}

Constructing an effective field theory for magnetically doped spin-orbit coupled semiconductors [Eq. (\ref{EFT_Free_Energy})], we performed the renormalization group analysis [Eqs. (\ref{eq:callan-symanzik}), (\ref{eq:beta functions}), and (\ref{eq:anomalousdimensions})] and obtained beta functions [Eq. (\ref{re:beta functions})] for all coupling parameters (the fermion-boson Zeeman interaction, the boson self-interaction, the disorder scattering, the fermion mass, the boson velocity, and the boson mass), evaluating Feynman diagrams up to the one-loop level [Eqs. (\ref{re:counter-fermion}), (\ref{re:counter-boson}), and (\ref{re:counter-vertex})]. Solving these renormalization group equations, we found an interacting fixed point [Eq. (\ref{re:rgsol1})] and revealed how all parameters evolve as a function of temperature near the fixed point [Eq. (\ref{re:rgsol2})]. In particular, we proposed a phase diagram in the plane of the applied magnetic field and temperature at a given concentration of magnetic impurities based on Eq. (\ref{re:field_wm}) [Fig. \ref{fig:phasediagram}]. In addition, we suggested the temperature $\&$ magnetic-field evolution for the longitudinal magnetoconductivity [Eqs. (\ref{re:cw}) and (\ref{re:LMC})] at low temperatures, where the Weyl metal phase is realized, given by Fig. \ref{fig:conductivity}.

Recently, we applied the present scenario to magnetically doped spin-orbit coupled semiconductors such as $Eu_{x}Bi_{2-x}Se_{3}$ and $Gd_{x}Bi_{2-x}Te_{3-y}Se_{y}$ \cite{Experiments_Doping_MI}. Actually, we could fit the temperature evolution for the enhancement factor $C_{W}$ of the longitudinal magnetoconductivity at low temperatures, where a Weyl metallic state is realized. Furthermore, experiments could extract out a phase diagram in the plane of the applied magnetic field and temperature at a given concentration of magnetic impurities, taking into account the maximum point of the longitudinal magnetoresistivity, which determines a critical magnetic field at a given temperature. It turns out that such a phase diagram is consistent with our proposed phase diagram.

One unsatisfactory point in our renormalization group analysis is that the regularization parameter $\ve$ was utilized as a phenomenological fitting parameter. Theoretically speaking, it should be chosen as $\ve \rightarrow 0$ in the last stage. However, the interacting fixed point turns out not to exist in the exactly three dimensional space. If we accept $\ve = 0.31$ literally, such an interacting fixed point can appear in quasi two dimensional systems. According to experiments \cite{Experiments_Doping_MI}, these semiconductors are rather anisotropic, where the in-plane resistivity ($\sim 0.001$ $\Omega$ $cm$) is two-order of magnitude smaller than the out-of-plane one ($\sim 0.1$ $\Omega$ $cm$), regarded to be quasi two dimensional. In addition, further doping of magnetic impurities is expected to make such magnetically doped semiconductors more isotropic, where the negative longitudinal magnetoresistivity disappears, indeed. It would be important to reveal the distribution pattern of doped magnetic impurities clearly in order to understand the physical meaning of the finiteness of the regularization parameter for our interacting fixed point.

\section*{ACKNOWLEDGEMENT}

This study was supported by the Ministry of Education, Science, and Technology (No. NRF-2015R1C1A1A01051629 and No. 2011-0030046) of the National Research Foundation of Korea (NRF).

\appendix

\begin{widetext}

\section{Evaluation of Feynman diagrams} \label{app:eval-diagrams}

\subsection{Fermion self-energy} \label{app:eval-fermionself}

The fermion self-energy (1) in Fig. \ref{fig:self-energy} is
\bea\ds\Sigma(1)&=&\sum_{i=1}^{d}\int\f{d^{d+\et}l}{(2\pi)^{d+\et}}(\lambda\gamma_{i}\gamma_{5})G_{0}(k-l)(\lambda\gamma_{i}\gamma_{5})D_{0}(l)\non
&=&\lambda^{2}\sum_{i}\int\f{d^{d+\et}l}{(2\pi)^{d+\et}}\f{\gamma_{i}\gamma_{5}[-\i(\k_{\tau}-\l_{\tau})\cdot\bg_{\tau}+(\k-\l)\cdot\bg-m]\gamma_{i}\gamma_{5}}{[(\k_{\tau}-\l_{\tau})^{2}+(\k-\l)^{2}+m^{2}][\l_{\tau}^{2}+c^{2}\l^{2}+r]}. \label{eq:fermion self}\eea
This can be evaluated in the standard way: Feynman parametrization, a momentum shift, and a spherically-symmetric integration as
\bea\Sigma(1)&=&\lambda^{2}\int^{1}_{0}dx\int\f{d^{d+\et}l}{(2\pi)^{d+\et}}\f{-\i d(\k_{\tau}-\l_{\tau})\cdot\bg_{\tau}+(d-2)(\k-\l)\cdot\bg-dm}{[(\l_{\tau}-x\k_{\tau})^{2}+a_{x}^{2}(\l-x\k/a_{x}^{2})^{2}+\Delta]}\non
&=&\lambda^{2}\int^{1}_{0}dx\int\f{d^{d+\et}\tilde{l}}{(2\pi)^{d+\et}}\f{-\i d(1-x)\k_{\tau}\cdot\bg_{\tau}+(d-2)(1-x)\f{c^{2}}{a_{x}^{2}}(\k\cdot\bg)-dm}{a_{x}^{d}[\tilde{l}^{2}+\Delta]^{2}}\non
&=&\f{\lambda^{2}\Gamma(\f{4-d-\et}{2})}{(4\pi)^{(d+\et)/2}}\int^{1}_{0}dx\f{-\i d(1-x)\k_{\tau}\cdot\bg_{\tau}+(d-2)(1-x)\f{c^{2}}{a_{x}^{2}}(\k\cdot\bg)-dm}{a_{x}^{d}[\Delta]^{\f{4-d-\et}{2}}},\eea
where $\Delta=x(1-x)(\k_{\tau}^{2}+c^{2}\k^{2}/a_{x}^{2})+xm^{2}+(1-x)r$ and $a_{x}^{2}=x+(1-x)c^{2}$. In the second line, the loop momentum is scaled as $\l\rightarrow\l/a_{x}$ and then redefined as $\tilde{l}=(\l_{\tau}-x\k_{\tau},\l-x\k/a_{x})$.
Near $d=3$ and $\et=1$, we get
\be\Sigma(1)=\f{\lambda^{2}}{8\pi^{2}\ve}[-\i3n_{1}\k_{\tau}\cdot\bg_{\tau}+n_{2}\k\cdot\bg-3n_{3}m]+\O(1),\ee
where
\be n_{1}=\int^{1}_{0}dx\f{1-x}{a_{x}^{3/2}}=\f{2}{c(1+c)^{2}},~n_{2}=\int^{1}_{0}dx\f{(1-x)c^{2}}{a_{x}^{5/2}}=\f{2(1+2c)}{3c(1+c)^{2}},~n_{3}=\int^{1}_{0}dx\f{1}{a_{x}^{3/2}}=\f{2}{c(1+c)}.\ee

Using Eq. (\ref{eq:counter}), we find the counterterms for the fermion dynamics as
\be\delta_{0}=-\f{3\lambda^{2}}{4\pi^{2}\ve c(1+c)^{2}},~~\delta_{1}=-\f{(1+2c)\lambda^{2}}{12\pi^{2}\ve c(1+c)^{2}},~~\delta_{m}=+\f{3\lambda^{2}}{4\pi^{2}\ve c(1+c)}.\label{re:fermionself}\ee
This result is comparable with that of QED in four dimensions, referred to as QED4. In QED4, we have $\delta_{\psi}=\f{e^{2}(2-d)}{16\pi^{2}(4-d)}\overset{d\rightarrow4}{\longrightarrow}-\f{e^{2}}{8\pi^{2}\ve}$ and $\delta_{m}=\f{e^{2}(-d)}{8\pi^{2}(4-d)}\overset{d\rightarrow4}{\longrightarrow}-\f{e^{2}}{2\pi^{2}\ve}$. Setting $c=1$ and tracking the dimensional factors, we observe that the above counterterms are reduced into
\be\delta_{0}\overset{d\rightarrow3}{\longrightarrow}-\f{3\lambda^{2}}{16\pi^{2}\ve},~~\delta_{1}\overset{d\rightarrow3}{\longrightarrow}-\f{\lambda^{2}}{16\pi^{2}\ve},~~\delta_{m}\overset{d\rightarrow3}{\longrightarrow}+\f{3e^{2}}{8\pi^{2}\ve}.\ee
The $\bm{\Phi}$-field has three space components but without the time component, so it gives $3/2$, $1/2$, and $- 3/4$ factors for the field, the velocity, and the mass counterterm, respectively.

\subsection{Boson self-energy} \label{app:eval-bosonself}

In Fig. \ref{fig:self-energy}, bosons get self-energy corrections (2) from the Zeeman coupling, (3) and (4) from the boson self-interaction, (5) from the disorder scattering, respectively.

\subsubsection{Zeeman coupling}

The correction (2) is
\bea\Pi_{ii}(2)&=&-\lambda^{2}\int\f{d^{d+\et}l}{(2\pi)^{d+\et}}\tr[\gamma_{i}\gamma_{5}G_{0}(l+q)\gamma_{i}\gamma_{5}G_{0}(l)]\non
&=&-\lambda^{2}\int\f{d^{d+\et}l}{(2\pi)^{d+\et}}\f{4\big(-l\cdot(l+q)+2l_{i}(l_{i}+q_{i})+m^{2}\big)}{\big[(l+q)^{2}+m^{2}\big]\big[l^{2}+m^{2}\big]}\non
&=&-4\lambda^{2}\int^{1}_{0}dx\int\f{d^{d+\et}l}{(2\pi)^{d+\et}}\f{\f{-d-\et+2}{d+\et}(l+xq)^{2}+x(1-x)(q^{2}-2q_{i}^{2})+m^{2}}{[(l+xq)^{2}+x(1-x)q^{2}+m^{2}]^{2}}\non
&=&-\f{4\lambda^{2}}{(4\pi)^{(d+\et)/2}}\int^{1}_{0}dx\f{(2x(1-x)q^{2}-2x(1-x)q_{i}^{2}+2m^{2})\Gamma(\f{4-d-\et}{2})}{[x(1-x)q^{2}+m^{2}]^{\f{4-d-\et}{2}}}.\label{eq:boson self}\eea
Near $d=3$ and $\et=1$, we get
\be\Pi_{ii}(2)=-\f{\lambda^{2}(\q_{\tau}^{2}+\q^{2}-\q_{i}^{2})}{6\pi^{2}\ve}-\f{\lambda^{2}m^{2}}{\pi^{2}\ve}.\label{re:boson self zeeman}\ee

\subsubsection{Boson interaction}

The correction (3) is
\be\Pi_{ii}(3)=-\f{Nu}{2}\int\f{d^{d+\et}l}{(2\pi)^{d+\et}}D_{0}(q-l)=-\f{Nu}{2}\int\f{d^{d+\et}l}{(2\pi)^{d+\et}}\f{1}{\l_{\tau}^{2}+c^{2}\l^{2}+r}=-\f{Nu\Gamma(\f{2-d-\et}{2})}{2(4\pi)^{(d+\et)/2}c^{d}r^{\f{2-d-\et}{2}}}.\ee
Near $d=3$ and $\ve_{\tau}=1$, we get $\Pi_{ii}(3)=\f{Nur}{16\pi^{2}\ve c^{3}}$. The correction (4) in the same line is similar except for the absence of $\f{N}{2}$ factor. The result is $\Pi_{ii}(4)=\f{ur}{8\pi^{2}\ve c^{3}}$.

\subsubsection{Disorder scattering}

The correction (5) is
\be\Pi_{ii}(5)=\Gamma_{m}\int\f{d^{d+\et}l}{(2\pi)^{d}}D_{0}(q-l)\delta^{(\et)}(\l_{\tau})=\Gamma_{m}\int\f{d^{d}\l}{(2\pi)^{d}}\f{1}{c^{2}\l^{2}+\q_{\tau}^{2}+r}=\f{\Gamma_{m}\Gamma(\f{2-d}{2})}{(4\pi)^{d/2}c^{d}(\q_{\tau}^{2}+r)^{\f{2-d}{2}}}.\ee
Near $d=3$ and $\et=1$, we get $\Pi_{ii}(5)=-\f{4\Gamma_{m}(\q_{\tau}^{2}+r)}{(4\pi)^{3/2}(\ve+\et)c^{3}}$.

\subsubsection{Counterterms}

The total boson self-energy is
\bea\Pi_{ii}(q)&=&\Pi_{ii}(2)+\Pi_{ii}(3)+\Pi_{ii}(4)+\Pi_{ii}(5)\non
&=&-\f{\lambda^{2}(\q_{\tau}^{2}+\q^{2}-\q_{i}^{2})}{6\pi^{2}\ve}-\f{\lambda^{2}m^{2}}{\pi^{2}\ve}+\f{(N+2)ur}{16\pi^{2}\ve c^{3}}-\f{4\Gamma_{m}(\q_{\tau}^{2}+r)}{(4\pi)^{3/2}(\ve+\et)c^{3}}.\eea
The Zeeman coupling term results in two unusual features and their corresponding complications. First, it has a transverse-mode structure ($\Pi_{ii}\sim\q^{2}-\q_{i}^{2}$) while the original action doesn't. This is not surprising because bosons are coupled to Dirac fermions in the fashion of the ``gauge field" (chiral).  However, we'll just ignore $\q_{i}^{2}$ to find the counterterms by assuming that the effective action is ``chosen" for the Feynman gauge so that no projection appears in Eq. (\ref{eq:action}). Second, a mass-shift proportional to the fermion mass appears. This makes the boson mass increase even at the critical point. Thus, in a naive view point, there is no critical state. Recall that the similar thing happens in the $\phi^{4}$-theory with a cutoff ($\Lambda$), where a mass shift proportional to $\Lambda^{2}$ appears. Redefinition of the boson mass with including the shift is required to access the critical point. Practically, this can be done by eliminating the additional mass shift. We'll do the same thing here: the mass shift is canceled by an ad hoc counterterm, not participating in the renormalization group.

Based on the above discussion, we find the counterterms for the bosons as
\be\delta_{2}=-\f{\lambda^{2}}{6\pi^{2}\ve}-\f{4\Gamma_{m}}{(4\pi)^{3/2}(\ve+\et)c^{3}},~\delta_{c}=-\f{\lambda^{2}}{6\pi^{2}\ve c^{2}},~\delta_{r}=\f{(N+2)u}{16\pi^{2}\ve c^{3}}-\f{4\Gamma_{m}}{(4\pi)^{3/2}(\ve+\et)c^{3}}.\ee
Compared with the result of QED4, $\delta_{A}=-\f{e^{2}}{6\pi^{2}\ve}$, the counterterms have the same numerical factor. However, due to the presence of velocity factors, $c$ gets renormalized for the fermion velocity $v=1$. Actually, this renormalization structure appears ubiquitously \cite{velocityRG} at least in the models that critical bosons are coupled to Dirac fermions.

\subsection{Vertex corrections} \label{app:eval-vertex corrections}

The boson self-interaction and the disorder scattering can be mixed, while the Zeeman coupling gets affected from the others only through the field renormalization. We calculate the disorder scattering, the boson interaction, mixing of them, and the Zeeman coupling in turn.

\subsubsection{Disorder scattering}

We calculate the diagrams of (1), (2), and (3) in Fig. \ref{fig:vertexcorrections}. The correction (1) is
\bea\delta\Gamma_{m}(1)&=&\Gamma_{m}^{2}\int\f{d^{d+\et}l}{(2\pi)^{d+\et}}D_{0}(k+l)D_{0}(k'-l)(2\pi)^{\et}\delta^{(\et)}(\l_{\tau})\non
&=&\f{\Gamma_{m}^{2}}{c^{d}}\int^{1}_{0}dx\int\f{d^{d+\et}l}{(2\pi)^{d+\et}}\f{(2\pi)^{\et}\delta^{(\et)}(\l_{\tau})}{[(l+xk+(1-x)k')^{2}+x(1-x)(k-k')^{2}+r]^{2}}\non
&=&\f{\Gamma_{m}^{2}}{c^{d}}\int^{1}_{0}dx\int\f{d^{d}\l}{(2\pi)^{d}}\f{1}{[(\l+x\k+(1-x)\k')^{2}+\Delta]^{2}}\non
&=&\f{\Gamma_{m}^{2}}{(4\pi)^{d/2}c^{d}}\int^{1}_{0}dx\f{\Gamma(\f{4-d}{2})}{[\Delta]^{\f{4-d}{2}}},\eea
where $\Delta=(x\k_{\tau}+(1-x)\k'_{\tau})^{2}+x(1-x)(k-k')^{2}+r$. Near $d=3$ and $\ve_{\tau}=1$, we find $\delta\Gamma_{m}(1)=\f{2\Gamma_{m}^{2}}{(4\pi)^{3/2}c^{3}(\ve+\et)}$. Other corrections differ from this only with numerical factors. The correction (2) is the same. The correction (3) is two times larger because there are two inequivalent diagrams. As a result, we find
\be\delta\Gamma_{m}(1)=\f{2\Gamma_{m}^{2}}{(4\pi)^{3/2}c^{3}(\ve+\et)},~\delta\Gamma_{m}(2)=\f{2\Gamma_{m}^{2}}{(4\pi)^{3/2}c^{3}(\ve+\et)},~\delta\Gamma_{m}(3)=\f{4\Gamma_{m}^{2}}{(4\pi)^{3/2}c^{4}(\ve+\et)}.\label{re:vc1}\ee

\subsubsection{Boson interaction}

We calculate the diagrams of (6), (7), (8), and (9) in Fig. \ref{fig:vertexcorrections}. The correction (6) is
\bea\delta u(6)&=&u^{2}\int\f{d^{d+\et}l}{(2\pi)^{d+\et}}D_{0}(q+l)D_{0}(l)=\f{u^{2}}{c^{d}}\int\f{d^{d+\et}l}{(2\pi)^{d+\et}}\f{1}{((q+l)^{2}+r)(l^{2}+r)}\non
&=&\f{u^{2}}{c^{d}}\int^{1}_{0}dx\int\f{d^{d+\et}l}{(2\pi)^{d+\et}}\f{1}{[(l+xq)^{2}+\Delta]^{2}}=\f{u^{2}}{(4\pi)^{\f{d+\et}{2}}c^{d}}\int^{1}_{0}dx\f{\Gamma(\f{4-d-\et}{2})}{[\Delta]^{\f{4-d-\et}{2}}},\eea
where $\Delta=x(1-x)q^{2}+r$. We find $\delta u(6)=\f{u^{2}}{8\pi^{2}\ve c^{3}}$. The other corrections can be found similarly. The correction (7) is the same. The correction (8) is two times larger because there are two inequivalent diagrams. The correction (9) is $\f{N}{2}$ times the first correction, where $2$ comes from two equivalent vertices and $N$ results from a free index summation. As a result, we find
\be\delta u(6)=\f{u^{2}}{8\pi^{2}\ve c^{3}},~~\delta u(7)=\f{u^{2}}{8\pi^{2}c^{3}\ve},~~\delta u(8)=\f{u^{2}}{4\pi^{2}c^{3}\ve},~~\delta u(9)=\f{Nu^{2}}{16\pi^{2}c^{3}\ve}.\label{re:vc2}\ee

\subsubsection{Mixing of boson interaction and disorder scattering}

The disorder scattering gets corrections from the diagrams of (4) and (5) in Fig. \ref{fig:vertexcorrections}. The calculation of (4) is similar with that of (1) because frequency is not exchanged in the loop. Numerical factor is $2N$, where $2$ comes from two inequivalent diagrams and $N$ results from a free index summation. The calculation of (5) is similar with that of (6) because frequency is exchanged in this case. Numerical factor is $2$ because there are two inequivalent diagrams. As a result, we find
\be\delta\Gamma_{m}(4)=-\f{4Nu\Gamma_{m}}{(4\pi)^{3/2}c^{3}(\ve+\et)},~\delta\Gamma_{m}(5)=-\f{u\Gamma_{m}}{4\pi^{2}c^{3}\ve}.\label{re:vc3}\ee
The boson interaction gets corrections from the diagrams of (10), (11), and (12). The calculations are similar with that of (1) because frequency is not exchanged. Numerical factors are $2$ because there are two inequivalent diagrams. As a result, we find
\be\delta u(10)=-\f{4u\Gamma_{m}}{(4\pi)^{3/2}c^{3}(\ve+\et)},~\delta u(11)=-\f{4u\Gamma_{m}}{(4\pi)^{3/2}c^{3}(\ve+\et)},~\delta u(12)=-\f{4u\Gamma_{m}}{(4\pi)^{3/2}c^{3}(\ve+\et)}.\label{re:vc4}\ee

\subsubsection{Zeeman coupling term}

The correction (14) in Fig. \ref{fig:vertexcorrections} is
\bea\delta\lambda(14)\gamma_{i}\gamma_{5}&=&\sum_{j=1}^{3}\int\f{d^{d+\et}l}{(2\pi)^{d+\et}}D_{0}(p-l)(\lambda\gamma_{j}\gamma_{5})G_{0}(l+q)(\lambda\gamma_{i}\gamma_{5})G_{0}(l)(\lambda\gamma_{j}\gamma_{5})\non
&=&\lambda^{3}\sum_{j}\int\f{d^{d+\et}l}{(2\pi)^{d+\et}}\f{\gamma_{j}\gamma_{5}[-\i(\l_{\tau}+\q_{\tau})\cdot\bg_{\tau}+(\l+\q)\cdot\bg-m]\gamma_{i}\gamma_{5}[-\i\l_{\tau}\cdot\bg_{\tau}+\l\cdot\bg-m]\gamma_{j}\gamma_{5}}{[(\p_{\tau}-\l_{\tau})^{2}+c^{2}(\p-\l)^{2}+r][(\l_{\tau}+\q_{\tau})^{2}+(\l+\q)^{2}+m^{2}][\l_{\tau}^{2}+\l^{2}+m^{2}]}.\eea
This is rearranged as
\be\delta\lambda(14)\gamma_{i}\gamma_{5}=\lambda^{3}\int^{1}_{0}dxdydz\delta(1-x-y-z)\int\f{d^{d+\et}\tilde{l}}{(2\pi)^{d+\et}}\f{2\N}{\D^{3}},\ee
where
\bea\D&=&\tilde{\l}_{\tau}^{2}+a_{x}^{2}\tilde{\l}^{2}+\Delta,\non
\tilde{l}&=&(\l_{\tau}-x\p_{\tau}+y\q_{\tau},\l-(xc^{2}/a_{x}^{2})\p+(y/a_{x}^{2})\q),\non
\Delta&=&x(1-x)\p_{\tau}^{2}+y(1-y)\q_{\tau}^{2}+2xy\p_{\tau}\cdot\q_{\tau}+\big(xyc^{2}(\p+\q)^{2}+xzc^{2}\p^{2}+yz\q^{2}\big)/a_{x}^{2}+xr+(1-x)m^{2},\non
a_{x}^{2}&=&xc^{2}+(1-x).\label{eq:deno}\eea
The numerator is given by
\bea\N&=&\sum_{j}\gamma_{j}\gamma_{5}[-\i(\l_{\tau}+\q_{\tau})\cdot\bg_{\tau}+(\l+\q)\cdot\bg-m]\gamma_{i}\gamma_{5}[\l_{\tau}\cdot\bg_{\tau}+\l\cdot\bg-m]\gamma_{j}\gamma_{5}\non
&=&\sum_{j}\gamma_{j}\gamma_{5}[-\i\tilde{\l}_{\tau}\cdot\bg_{\tau}+\tilde{\l}\cdot\bg]\gamma_{i}\gamma_{5}[-\i\tilde{\l}_{\tau}\cdot\bg_{\tau}+\tilde{\l}\cdot\bg]\gamma_{j}\gamma_{5}+\O(\tilde{l})\non
&\rightarrow&[(d-2)\tilde{\l}_{\tau}^{2}+(d-2)^{2}\tilde{l}_{k}^{2}]\gamma_{i}\gamma_{5}.\eea
In the last line, only the quadratic terms are kept. The integration is straightforward, and the result is
\be\delta\lambda(14)=\lambda^{3}\int^{1}_{0}dxdydz\delta(1-x-y-z)\f{\Gamma(\f{4-d-\et}{2})[(d-2)+(d-2)^{2}/a_{x}^{2}]}{a_{x}^{d}2(4\pi)^{(d+1)/2}\Delta^{\f{4-d-\et}{2}}}.\ee
Near $d=3$ and $\ve_{\tau}=1$, we find
\be\delta\lambda(14)=\f{\lambda^{3}}{16\pi^{2}\ve}\int^{1}_{0}dxdydz\delta(1-x-y-z)(a_{x}^{-3}+a_{x}^{-5})=\f{\lambda^{3}(1+2c)}{12\pi^{2}\ve c(1+c)^{2}}.\label{re:vc5}\ee

Lastly, we calculate the correction (13). It should be interpreted as an amputated diagram since six distinct diagrams appear when it is contracted to external lines. It possibly gives a correction for the boson interaction because a power counting tells that it may have a logarithmic divergence. However, it turns out not to diverge. The correction (13) is
\bea\delta u(13)&=&-\lambda^{4}\int\f{d^{d+\et}l}{(2\pi)^{d+\et}}\tr[\gamma_{i}\gamma_{5}G_{0}(l)\gamma_{j}\gamma_{5}G_{0}(l+k)\gamma_{k}\gamma_{5}G_{0}(l-q)\gamma_{l}\gamma_{5}G_{0}(l-k')]\non
&=&-\lambda^{4}\int^{1}_{0}dxdydzdw\int\f{d^{d+\et}l}{(2\pi)^{d+\et}}\delta(1-x-y-z-w)\f{3!\N}{\D^{4}}.\eea
The denominator and the numerator are
\bea\D&=&(l+yk-zq-wk')^{2}+\Delta,\non
\Delta&=&y(1-y)k^{2}+z(1-z)q^{2}+w(1-w)k^{'2}+2(yzk\cdot q-zwq\cdot k'+ywk\cdot k'),\non
\N&=&\tr[\gamma_{i}\gamma_{5}(l\cdot\gamma)\gamma_{j}\gamma_{5}((l+k)\cdot\gamma)\gamma_{k}\gamma_{5}((l-q)\cdot\gamma)\gamma_{l}\gamma_{5}((l-k')\cdot\gamma)]\non
&\rightarrow&\tr[\gamma_{i}\gamma_{5}\tilde{l}\cdot\gamma\gamma_{j}\gamma_{5}\tilde{l}\cdot\gamma\gamma_{k}\gamma_{5}\tilde{l}\cdot\gamma\gamma_{l}\gamma_{5}\tilde{l}\cdot\gamma],\eea
where $\tilde{l}=l+yk-zq-wk'$ and only the quartic term is kept in the last line. The integration is straightforward and the result is
\bea\delta u(13)&=&-\f{\lambda^{4}\Gamma(\f{4-d-\et}{2})}{(4\pi)^{(d+\et)/2}}\int^{1}_{0}dxdydzdw\f{\delta(1-x-y-z-w)}{[\Delta]^{\f{4-d-\et}{2}}}\non
&&\times\f{1}{4}(\delta_{\mu\nu}\delta_{\rho\sigma}+\delta_{\mu\rho}\delta_{\nu\sigma}+\delta_{\mu\sigma}\delta_{\nu\rho})\tr[\gamma_{i}\gamma_{5}\gamma_{\mu}\gamma_{j}\gamma_{5}\gamma_{\nu}\gamma_{k}\gamma_{5}\gamma_{\rho}\gamma_{l}\gamma_{5}\gamma_{\sigma}].\eea
Near $d=3$ and $\et=1$, we find
\be\delta u(13)=-\f{\lambda^{4}}{192\pi^{2}\ve}(\delta_{\mu\nu}\delta_{\rho\sigma}+\delta_{\mu\rho}\delta_{\nu\sigma}+\delta_{\mu\sigma}\delta_{\nu\rho})\tr[\gamma_{i}\gamma_{\mu}\gamma_{j}\gamma_{\nu}\gamma_{k}\gamma_{\rho}\gamma_{l}\gamma_{\sigma}],\ee
where we used $\int^{1}_{0}dxdydzdw\delta(1-x-y-z-w)=\f{1}{6}$. Taking the trace, we have
\bea&&\tr[\gamma_{i}\gamma_{\mu}\gamma_{j}\gamma_{\mu}\gamma_{k}\gamma_{\nu}\gamma_{l}\gamma_{\nu}]=\tr[\gamma_{i}\gamma_{\mu}\gamma_{j}\gamma_{\nu}\gamma_{k}\gamma_{\nu}\gamma_{l}\gamma_{\mu}]\simeq16(\delta_{ij}\delta_{kl}-\delta_{ik}\delta_{jl}+\delta_{il}\delta_{jk}),\non
&&\tr[\gamma_{i}\gamma_{\mu}\gamma_{j}\gamma_{\nu}\gamma_{k}\gamma_{\mu}\gamma_{l}\gamma_{\nu}]=-32\delta_{ik}\delta_{jl},\eea
where $\O(\ve)$ terms are dropped. In the calculation, we used the following identities repeatedly:
\bea\gamma_{\mu}\gamma_{\nu}\gamma_{\mu}&=&(d-2+\et)\gamma_{\nu},\non
\gamma_{\mu}\gamma_{\nu}\gamma_{\rho}\gamma_{\mu}&=&4\delta_{\nu\rho}-(d-4+\et)\gamma_{\nu}\gamma_{\rho},\non
\gamma_{\mu}\gamma_{\nu}\gamma_{\rho}\gamma_{\sigma}\gamma_{\mu}&=&2\gamma_{\sigma}\gamma_{\rho}\gamma_{\nu}+(d-4+\et)\gamma_{\nu}\gamma_{\rho}\gamma_{\sigma}.\eea
As a result, we find
\be\delta u(13)=-\f{\lambda^{4}}{6\pi^{2}\ve}(\delta_{ij}\delta_{kl}-2\delta_{ik}\delta_{jl}+\delta_{il}\delta_{jk}).\ee
Contracted to external fields, each gives rise to two diagrams because there are four options but each two are the same. See Fig. \ref{fig:int-from-zeeman}. As a result, we obtain
\be\M_{1}=\M_{2}=\M_{3}=\M_{4}=-\f{\lambda^{4}}{6\pi^{2}\ve},~\M_{5}=\M_{6}=\f{\lambda^{4}}{3\pi^{2}\ve}.\ee
Note that the sum is zero $\sum_{i=1}^{6}\M_{i}=0$. Thus, the logarithmic divergence is cancelled as well known in the Ward identity of QED4.

\bf[t]\centering\ing[width=0.5\tw]{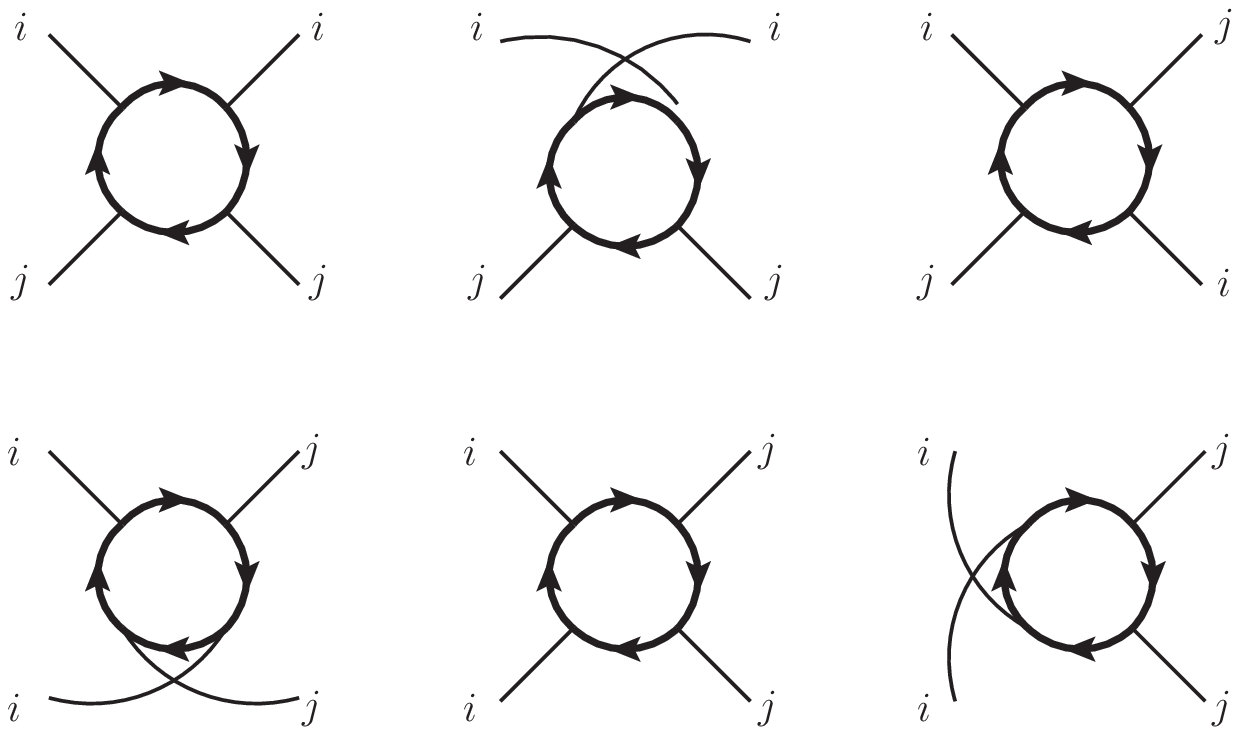}\caption{Vertex corrections for the boson interaction from the four Zeeman couplings. Three diagrams in the first line are $\M_{1}$, $\M_{2}$, and $\M_{3}$ from left to right, and three diagrams in the second, $\M_{4}$, $\M_{5}$, and $\M_{6}$.}\label{fig:int-from-zeeman}\ef

\subsubsection{Counterterms}

We gather all the results from Eqs. (\ref{re:vc1}), (\ref{re:vc2}), (\ref{re:vc3}), (\ref{re:vc4}), and (\ref{re:vc5}) as
\bea&&\sum_{i=1}^{5}\delta\Gamma_{m}(i)=\f{8\Gamma_{m}^{2}-4Nu\Gamma_{m}}{(4\pi)^{3/2}c^{3}(\ve+\et)}-\f{u\Gamma_{m}}{4\pi^{2}c^{3}\ve}-\f{4Nu\Gamma_{m}}{(4\pi)^{3/2}c^{3}(\ve+\et)},\non
&&\sum_{i=6}^{12}\delta u(i)=\f{(N+8)u^{2}}{16\pi^{2}c^{3}\ve}-\f{12u\Gamma_{m}}{(4\pi)^{3/2}c^{3}(\ve+\et)},~~\sum_{i=14}^{14}\delta\lambda(i)=\f{\lambda^{3}(1+2c)}{12\pi^{2}c(1+c)^{2}\ve}.\eea

Using Eq. (\ref{eq:counter}), we find
\bea&&\delta_{\Gamma_{m}}=-\f{8\Gamma_{m}}{(4\pi)^{3/2}c^{3}(\ve+\et)}+\f{u}{4\pi^{2}c^{3}\ve}+\f{4Nu}{(4\pi)^{3/2}c^{3}(\ve+\et)},\non
&&\delta_{u}=\f{(N+8)u}{16\pi^{2}c^{3}\ve}-\f{12\Gamma_{m}}{(4\pi)^{3/2}c^{3}(\ve+\et)},~~\delta_{\lambda}=-\f{\lambda^{2}(1+2c)}{12\pi^{2}c(1+c)^{2}\ve}.\eea

\section{Fixed point structure}\label{app:fixed point}

The beta functions are given by
\bea\beta_{c}&=&c\big[-f_{1}(c)\tilde{\alpha}+\tilde{\Gamma}\big],\non
\beta_{\tilde{\alpha}}&=&\tilde{\alpha}\big[-\ve+f_{2}(c)\tilde{\alpha}+\tilde{\Gamma}\big],\non
\beta_{\tilde{u}}&=&\tilde{u}\big[-\ve+11\tilde{u}+f_{3}(c)\tilde{\alpha}-5\tilde{\Gamma}\big],\non
\beta_{\tilde{\Gamma}}&=&\tilde{\Gamma}\big[-1-\ve-3\tilde{\Gamma}+f_{4}(c)\tilde{\alpha}+4(6\sqrt{\pi}+1)\tilde{u}\big],\non
\beta_{m}&=&m\big[-1+f_{5}(c)\tilde{\alpha}\big],\non
\beta_{r}&=&r\big[-2+f_{6}(c)\tilde{\alpha}+5\tilde{u}\big],\eea
where $\tilde{\alpha}$, $\tilde{u}$, and $\tilde{\Gamma}$ are defined in Eq. (\ref{eq:dimensionless couplings}) and
\bea&&f_{1}(c)=-\f{c^{4}+2c^{3}+2c^{2}-10c-1}{c(1+c)^{2}},~f_{2}(c)=\f{c^{2}+1}{c},~f_{3}(c)=\f{3+c^{2}}{c},\\
&&f_{4}(c)=\f{c^{4}+2c^{3}+6c^{2}-2c+3}{c(1+c)^{2}},~f_{5}(c)=\f{10+11c}{(1+c)^{2}},~f_{6}(c)=\f{2(c^{3}+2c^{2}+3c-8)}{(1+c)^{2}}.\nn\eea

There are two stable fixed points. The first fixed point (FP1) is
\be\tilde{\alpha}_{*}=0,~\tilde{u}_{*}=\f{5+2\ve}{120\sqrt{\pi}-13},~\tilde{\Gamma}_{*}=\f{11+\ve(7-24\sqrt{\pi})}{120\sqrt{\pi}-13},~c_{*}=0,\label{re:fp1}\ee
where it is stable when $0\le\ve<0.08318$. All coupling constants are zero but the ratios of the boson interaction and the disorder strength to the boson velocity are finite. This results from the fact that the screening effect of the disorder strength is so strong ($\tilde{\Gamma}_{*}>\ve/2$) that the Zeeman coupling cannot have a non-Gaussian fixed point ($\beta_{\tilde{\alpha}}>0$). The disorder strength makes the boson velocity decrease ($\beta_{c}\propto-c\tilde{\Gamma}_{*}$). Since the boson velocity goes to zero, the ratios of $\tilde{u}\sim u/c^{3}$ and $\tilde{\Gamma}\sim\Gamma_{m}/c^{3}$ are finite.

The anomalous dimensions are given by
\be z=1,~~\eta_{\psi}=0,~~\eta_{\Phi}=0.0551-0.178\ve,~~\nu_{m}^{-1}=1,~~\nu_{r}^{-1}=1.96-0.805\ve.\label{re:par_fp1}\ee
Note that the scaling law of fermions is trivial: $z=1$, $\eta_{\psi}=0$, and $\nu_{m}=1$. This is because fermions are non-interacting ($\lambda\rightarrow0$) in the low energy limit. On the other hand, the interaction strength and the disorder strength change the scaling law of bosons: $\eta_{\phi}\neq0$ and $\nu_{r}\neq0.5$. Thus, this phase is rather trivial in that it is thought to be a Wilson-Fisher fixed point with a finite disorder strength.

\begin{table}\begin{adjustbox}{max width=\textwidth}\centering\begin{tabular}{c|c|c|c|c}
$\tilde{\alpha}_{*}$&$\tilde{u}_{*}$&$\tilde{\Gamma}_{*}$&$c_{*}$&stability\\\hline
0&0&0&non-universal&unstable\\
0&$\f{\ve}{11}$&0&non-universal&unstable\\
0&0&$-\f{\ve+1}{3}$&0&unphysical (negative $\tilde{\Gamma}$)\\
0&$\f{5+2\ve}{120\sqrt{\pi}-13}$&$\f{11+\ve(7-24\sqrt{\pi})}{120\sqrt{\pi}-13}$&$0$&stable when $0\le\ve<0.08318$\\
$\f{\ve}{f_{2}}$&0&0&$f_{1}^{-1}(0)$&unphysical (stable when $\ve>4.055$)\\
$\f{\ve}{f_{2}}$&$\f{\ve(1-f_{3}/f_{2})}{11}<0$&0&$f_{1}^{-1}(0)$&unphysical (negative $\tilde{u}$)\\
$\f{4\ve+1}{3f_{2}+f_{4}}$&0&$\f{\ve f_{4}-(\ve+1)f_{2}}{3f_{2}+f_{4}}<0$&$[(4\ve+1)f_{1}+(\ve+1)f_{2}-\ve f_{4}]^{-1}(0)$&unphysical (negative $\tilde{\Gamma}$)\\
$\f{\ve}{f_{1}+f_{2}}$&$\f{\ve[6f_{1}+f_{2}-f_{3}]}{11[f_{1}+f_{2}]}$&$\f{\ve f_{1}}{f_{1}+f_{2}}$&$\big[\f{11(4f_{1}+f_{2}-f_{4})-4(6\sqrt{\pi}+1)[6f_{1}+f_{2}-f_{3}]}{11(f_{1}+f_{2})}\big]^{-1}(-\f{1}{\ve})$&exists and stable when $0.08318\le\ve<0.3103$
\end{tabular}\end{adjustbox}\caption{All fixed points in the one-loop order are shown.}\label{tab:fixedpoints}\end{table}

The second fixed point (FP2) is
\bea\ds&&\tilde{\alpha}_{*}=\f{\ve}{f_{1}(c_{*})+f_{2}(c_{*})},\tilde{u}_{*}=\f{\ve[6f_{1}(c_{*})+f_{2}(c_{*})-f_{3}(c_{*})]}{11[f_{1}(c_{*})+f_{2}(c_{*})]},\tilde{\Gamma}_{*}=\f{\ve f_{1}(c_{*})}{f_{1}(c_{*})+f_{2}(c_{*})},\nn\\
&&\f{4f_{1}(c_{*})+f_{2}(c_{*})-f_{4}(c_{*})}{f_{1}(c_{*})+f_{2}(c_{*})}-\f{4(6\sqrt{\pi}+1)}{11}\f{6f_{1}(c_{*})+f_{2}(c_{*})-f_{3}(c_{*})}{f_{1}(c_{*})+f_{2}(c_{*})}=-\f{1}{\ve},\label{re:fp2}\eea
where $c_{*}$ is obtained by solving the last equation. Numerically, the second fixed point is given by
\bea&&\tilde{\alpha}_{*}=6.897+2.446\ln\ve,~~\tilde{u}_{*}=4.052+0.7732\ve,\non
&&\tilde{\Gamma}_{*}=1.031+2.450\ve,~~c_{*}=1.367-0.02158\ve^{-1.75}.\label{re:par_fp2}\eea
We find that FP2 exists when $0.08318\le\ve$ because $\ve$ should be large enough to overcome the screening in $\beta_{\alpha}(\ve>2\tilde{\Gamma}$). It is stable only when $\ve<0.3103$. The anomalous dimensions are given by
\bea&&z=1.00+0.394\ve,~~\eta_{\psi}=-0.00427+0.346\ve,~~\eta_{\Phi}=0.5\ve,\non
&&\nu_{m}^{-1}=1.08-1.99\ve,~~\nu_{r}^{-1}=1.97-0.411\ve.\label{re:par_fp2_exps}\eea

The other fixed points can be found in Table \ref{tab:fixedpoints}. Most of them are either unphysical or unstable. The only stable fixed points are the fourth (FP1) and the last one (FP2), which are given in Eqs. (\ref{re:fp1}) and (\ref{re:par_fp2}), respectively. Stabilities of FP1 and FP2 can be checked with the linearized equations ($\Delta X\equiv X-X_{*}$):
\bea\fs\bpm\beta_{\Delta c}\\\beta_{\Delta\tilde{\alpha}}\\\beta_{\Delta\tilde{u}}\\\beta_{\Delta\tilde{\Gamma}}\epm&=&\bpm c_{*}&0&0&0\\0&\tilde{\alpha}_{*}&0&0\\0&0&\tilde{u}_{*}&0\\0&0&0&\tilde{\Gamma}_{*}\epm\bpm-\tilde{\alpha}_{*}f_{1}'(c_{*})&-f_{1}(c_{*})&0&1\\\tilde{\alpha}_{*}f_{2}'(c_{*})&f_{2}(c_{*})&0&1\\\tilde{\alpha}_{*}f_{3}'(c_{*})&f_{3}(c_{*})&11&-5\\\tilde{\alpha}_{*}f_{4}'(c_{*})&f_{4}(c_{*})&4(6\sqrt{\pi}+1)&-3\epm\bpm\Delta c\\\Delta\tilde{\alpha}\\\Delta\tilde{u}\\\Delta\tilde{\Gamma}\epm+\bpm\f{(\beta_{c})_{*}}{c_{*}}\Delta c\\\f{(\beta_{\tilde{\alpha}})_{*}}{\tilde{\alpha}_{*}}\Delta\tilde{\alpha}\\\f{(\beta_{\tilde{u}})_{*}}{\tilde{u}_{*}}\Delta\tilde{u}\\\f{(\beta_{\tilde{\Gamma}})_{*}}{\tilde{\Gamma}_{*}}\Delta\tilde{\Gamma}\epm.\label{eq:stability}\eea\ns
In Fig. \ref{fig:stability}, deviation from Eq. (\ref{re:fp1}) and that from Eq. (\ref{re:fp2}) go to zero in the low energy limit, so both FP1 and FP2 are infrared-stable.

\bf[t]\centering\ing[width=0.6\tw]{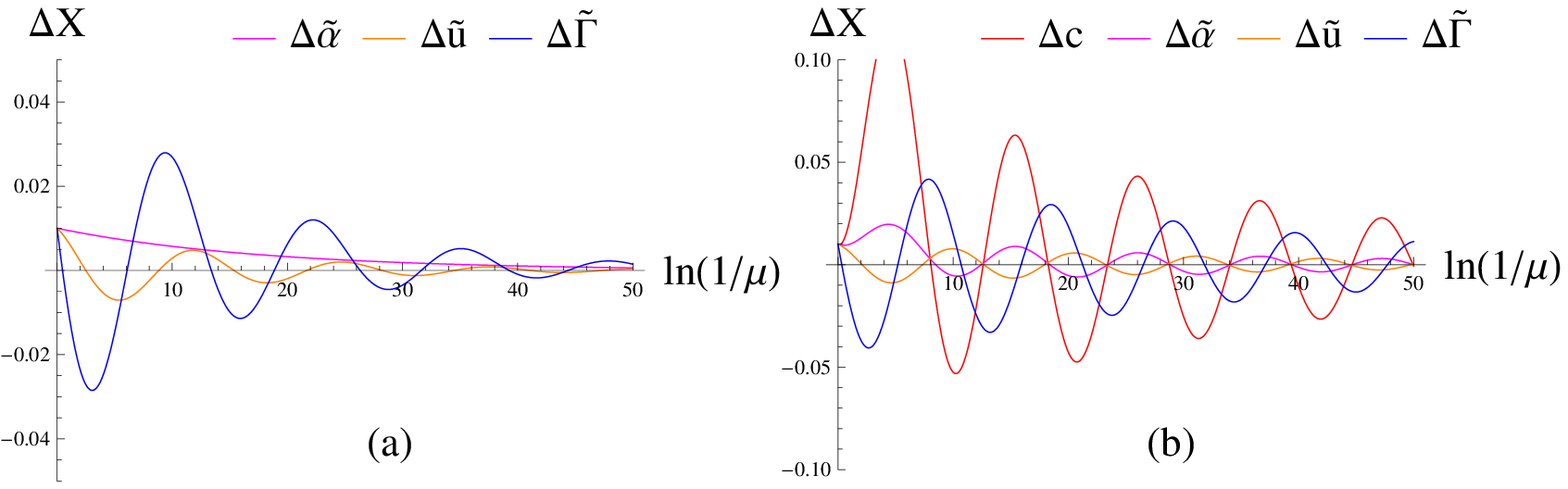}\caption{Flowing of deviations of coupling constants (a) from Eq. (\ref{re:fp1}) and (b) from Eq. (\ref{re:fp2})}\label{fig:stability}\ef

\section{fermi surface effect}\label{app:fermi surface}

In this section, we look into the consequence of a finite density of fermions on the boson dynamics. With a chemical potential $\mu$, the fermion action is given by
\be S_{f}=\int^{\beta}_{0}d\tau\int d^{d}\r\bar{\psi}\big((\partial_{\tau}-\mu)\gamma_{0}-\i\bg\cdot\partial_{\r}+m\big)\psi.\ee
Then, the boson self-energy in Eq. (\ref{eq:boson self}) is modified as
\be\Pi_{ii}(2)=4\lambda^{2}\int\f{d^{d+1}l}{(2\pi)^{4}}\f{(l_{0}+q_{0}-\i\mu)(l_{0}-\i\mu)+(\l+\q)\cdot\l-2(l_{i}+q_{i})l_{i}-m^{2}}{\big[(l_{0}+q_{0}-\i\mu)^{2}+{E'}^{2}\big]\big[(l_{0}-\i\mu)^{2}+E^{2}\big]},\ee
where $E'=\sqrt{(\l+\q)^{2}+m^{2}}$ and $E=\sqrt{\l^{2}+m^{2}}$.

Performing the integration over $l_{0}$, we have
\bea\Pi_{ii}'(2)&=&4\lambda^{2}\int\f{d^{d}\l}{(2\pi)^{d}} \theta(E'\geq|\mu|) \theta(E\geq|\mu|) \bigg\{\f{(E'+E)\big[E'E+(\l+\q)\cdot\l-2(l_{i}+q_{i})l_{i}-m^{2}\big]}{2E'E\big[q_{0}^{2}+(E'+E)^{2}\big]}\bigg\} , \non
\Pi''_{ii}(2)&=&4\lambda^{2}\int\f{d^{d}\l}{(2\pi)^{d}}\theta(E'\geq|\mu|)\theta(E\leq|\mu|)\bigg\{\f{(E'-E)\big[E'E-\big((\l+\q)\cdot\l-2(l_{i}+q_{i})l_{i}-m^{2}\big)\big]}{2E'E\big[q_{0}^{2}+(E'-E)^{2}\big]}\bigg\}\bigg],\eea
where $\Pi_{ii}(2)=\Pi_{ii}'(2)+\Pi_{ii}''(2)$ and $\theta(x)=1$ $(0)$ if $x$ is true (false).

First, we consider $\Pi_{ii}'(2)$. If $|\mu|\leq m$, it just goes back to the original expression in Eq. (\ref{eq:boson self}). On the other hand, if $|\mu|>m$, it changes as the chemical potential constrains the phase space with the theta function. Expanding $\Pi_{ii}'(2)$ with respect to the external momentum, we have
\bea\Pi_{ii}'(2)&=&-\f{\lambda^{2}(q_{0}^{2}+\q^{2}-q_{i}^{2})}{4}\int\f{d^{d}\l}{(2\pi)^{d}}\theta(|\l|\geq k_{F})\f{(\l^{2}-l_{i}^{2})}{E^{5}}+2\lambda^{2}\int\f{d^{d}\l}{(2\pi)^{d}}\theta(|\l|\geq k_{F})\f{(\l^{2}-l_{i}^{2})}{E^{3}}\non
&&+4\lambda^{2}\int\f{d^{d}\l}{(2\pi)^{d}}\delta(|\l|-k_{F}) \bigg\{-\f{(\l\cdot\q)^{2}}{4|\l|E^{3}}-\f{(\l\cdot\q)(l_{i}q_{i})}{2|\l|E^{3}} +\f{3(\l\cdot\q)^{2}(l_{i}^{2}+m^{2})}{4|\l|E^{5}}\bigg\},\eea
where $k_{F}=\sqrt{\mu^{2}-m^{2}}$. The first term gives a correction for the boson dynamics. There is a singular correction since the theta function does change a finite part but not a singular part. For the same reason, the second term gives a singular correction for the boson mass. The third term gives a finite correction for the $\q^{2}$-term since the delta function forces $\l$ onto $k_{F}$.
As a result, we obtain
\be\Pi_{ii}'(2)=-\f{\lambda^{2}(q_{0}^{2}+\q^{2}-q_{i}^{2})}{6\pi^{2}\ve}-\f{\lambda^{2}\mu^{2}}{\pi^{2}\ve}+\O(1),\ee
whose epsilon poles are the same with that of Eq. (\ref{re:boson self zeeman}) except for the change in the boson mass of $m^{2}\rightarrow\mu^{2}$.

Next, we consider $\Pi''(2)$ term. This term arises due to the finite density of fermions, where it vanishes in the case of $|\mu|<m$. It also vanishes in the limit of $q_{0}\rightarrow0$ so that we keep $q_{0}$ finite. Then, it is irregular with $\q$ (it cannot be expanded with respect to $|\q|$) as is in the standard Lindhard function. Keeping most singular terms, we have
\be\Pi_{ii}''(2)=4\lambda^{2}\int^{\infty}_{0}\f{dll^{d-1}}{(2\pi)^{3}}\int^{1}_{-\f{q}{2k_{F}}}dx\f{E\delta(l-k_{F})(lqx)^{2}}{l\big[(lqx)^{2}+q_{0}^{2}E^{2}\big]}\simeq\f{\lambda^{2}\mu k_{F}}{2\pi^{3}}-\f{\lambda^{2}\mu ^{2}}{4\pi^{2}}\f{|q_{0}|}{|\q|} .\ee
The first term gives a finite mass shift so it is not important. On the other hand, the second term, called the Landau damping term, changes the boson dynamics severely. In the low frequency regime, it is more singular than the quadratic term $q_{0}^{2}$, so this quadratic term cannot be used in this case.

Doing the similar calculation for Eq. (\ref{eq:fermion self}), one can show that the renormalization factors for the fermion dynamics are not changed. Also, the fermion dynamics does not get any singular corrections as opposed to the boson case. However, if we use the damped boson propagator ($\f{\lambda^{2}\mu^{2}}{4\pi^{2}}\f{|q_{0}|}{|\q|}+c^{2}\q^{2}$) instead of the bare one ($q_{0}^{2}+c^{2}\q^{2}$), then one may find that the fermion dynamics gets a singular correction with a logarithm factor ($\Sigma\sim\omega\log{\omega}$), well known to be a marginal Fermi liquid.

\section{Ward identity} \label{app:ward identity}

We prove the Ward identity in Eq. (\ref{eq:ward identity}). We focus on $S_{f}=\int dx\bar{\psi}(x)(\i\partial_{\mu}\gamma_{\mu}+m)\psi(x)$, where $\partial_{\mu}=(-\i\partial_{0},\partial_{\r})$ and $\gamma_{\mu}=(\gamma_{0},\bm{\gamma})$. Performing the chiral transformation given by
\be\psi(x)\rightarrow e^{\i\alpha(x)\gamma_{5}}\psi(x),\label{eq:chiral transformation-app}\ee
we have
\be\delta S_{f}=\int dx\{\partial_{\mu}\alpha(x)\bar{\psi}(x)\gamma_{\mu}\gamma_{5}\psi(x)+2\i m\alpha(x)\bar{\psi}(x)\gamma_{5}\psi(x)\}.\ee
Thus, the action is not invariant not only by the chiral current term but also by the pseudo-scalar term proportional to the mass.

Now, we calculate the fermion Green's function:
\be\left\langle\psi(x_{1})\bar{\psi}(x_{2})\right\rangle=\f{1}{\Z}\int\D(\bar{\psi},\psi)\psi(x_{1})\bar{\psi}(x_{2})e^{-S},\label{eq:fermion green-app}\ee
where $\Z = \int\D(\bar{\psi},\psi) e^{-S}$ is the partition function. If we assume Eq. (\ref{eq:fermion green-app}) is invariant under Eq. (\ref{eq:chiral transformation-app}) in spite of the explicit symmetry breaking, we have
\bea\ds0&=&-\int dx\left\langle\{\partial_{\mu}\alpha(x)\bar{\psi}(x)\gamma_{\mu}\gamma_{5}\psi(x)+2\i m\alpha(x)\bar{\psi}(x)\gamma_{5}\psi(x)\}\psi(x_{1})\bar{\psi}(x_{2})\right\rangle\nn\\
&&+\i\alpha(x_{1})\left\langle\gamma_{5}\psi(x_{1})\bar{\psi}(x_{2})\right\rangle+\i\alpha(x_{2})\left\langle\psi(x_{1})\bar{\psi}(x_{2})\gamma_{5}\right\rangle+\O(\alpha^{2})\nn\\
&=&\int dx\alpha(x)\bigr\{\partial_{\mu}\left\langle\bar{\psi}(x)\gamma_{\mu}\gamma_{5}\psi(x)\psi(x_{1})\bar{\psi}(x_{2})\right\rangle+2\i m\left\langle\bar{\psi}(x)\gamma_{5}\psi(x)\psi(x_{1})\bar{\psi}(x_{2})\right\rangle\nn\\
&&+\i\delta(x-x_{1})\left\langle\gamma_{5}\psi(x_{1})\bar{\psi}(x_{2})\right\rangle+\i\delta(x-x_{2})\left\langle\psi(x_{1})\bar{\psi}(x_{2})\gamma_{5}\right\rangle\bigl\}+\O(\alpha^{2}),\label{eq:invariance}\eea
where we performed integration by parts on $\partial_{\mu}\alpha(x)$ and used $\alpha(x_{1})=\int dx\delta(x-x_{1})\alpha(x)$.

The right hand side of Eq. (\ref{eq:invariance}) should be zero order by order. In the first order, we have
\bea&&\int dp_{1}\left\{k_{\mu}\left\langle\bar{\psi}(p_{1}+k)\gamma_{\mu}\gamma_{5}\psi(p_{1})\psi(q)\bar{\psi}(p)\right\rangle-2m\left\langle\bar{\psi}(p_{1}+k)\gamma_{5}\psi(p_{1})\psi(q)\bar{\psi}(p)\right\rangle\right\}\non&&~~~~~~~~~~~~~~~~~~=\left\langle\gamma_{5}\psi(q-k)\bar{\psi}(p)\right\rangle+\left\langle\psi(q)\bar{\psi}(p+k)\gamma_{5}\right\rangle.\eea
More compactly, we have
\be k_{\mu}G(p+k)\Gamma_{\mu 5}(p+k,p)G(p)-2mG(p+k)\Gamma_{5}(p+k,p)G(p)=\gamma_{5}G(p)+G(p+k)\gamma_{5},\ee
where $\Gamma_{\mu 5}(p+k,p)$ is the one-particle irreducible vertex for $\int dp_{1}\left\langle\bar{\psi}(p_{1}+k)\gamma_{\mu}\gamma_{5}\psi(p_{1})\psi(q)\bar{\psi}(p)\right\rangle$ and $\Gamma_{5}(p+k,p)$ is that for $\int dp_{1}\left\langle\bar{\psi}(p_{1}+k)\gamma_{5}\psi(p_{1})\psi(q)\bar{\psi}(p)\right\rangle$. Multiplying $G^{-1}(p+k)$ on the left and $G^{-1}(p)$ on the right, we obtain
\be k_{\mu}\Gamma_{\mu 5}(p+k,p)=G^{-1}(p+k)\gamma_{5}+\gamma_{5}G^{-1}(p)+2m\Gamma_{5}(p+k,p).\ee
Recalling $G^{-1}(p)=Z_{0}\i\p_{\tau}\cdot\bm{\gamma}_{\tau}-Z_{1}\p\cdot\bm{\gamma}-Z_{m}m$, we find $\gamma_{5}G^{-1}(p)=-G^{-1}(p)\gamma_{5}-2Z_{m}m\gamma_{5}$. Resorting to this expression, we obtain
\be Z_{\mu 5}\gamma_{\mu}\gamma_{5}=\f{\partial G^{-1}(p)}{\partial p_{\mu}}\gamma_{5}+2m(Z_{5}-Z_{m})\gamma_{5},\ee
where $Z_{\mu5}\gamma_{\mu}\gamma_{5}\equiv\lim_{k\rightarrow0}\Gamma_{\mu 5}(p+k,p)$ and $Z_{5}\gamma_{5}\equiv\lim_{k\rightarrow0}\Gamma_{5}(p+k,p)$. If $Z_{5}=Z_{m}$, we have the Ward identity:
\be Z_{\mu5}\gamma_{\mu}\gamma_{5}\overset{!}{=}\f{\partial G^{-1}(p)}{\partial p_{\mu}}\gamma_{5}.\label{eq:wardidentity}\ee

In order to verify Eq. (\ref{eq:wardidentity}), we calculate $Z_{5}$. It is given by
\bea\delta\Gamma_{5}\gamma_{5}&=&\sum_{j=1}^{3}\int\f{d^{d+\et}l}{(2\pi)^{d+\et}}D_{0}(p-l)(\lambda\gamma_{j}\gamma_{5})G_{0}(l+q)(\gamma_{5})G_{0}(l)(\lambda\gamma_{j}\gamma_{5})\non
&=&\lambda^{2}\sum_{j}\int\f{d^{d+\et}l}{(2\pi)^{d+\et}}\f{\gamma_{j}\gamma_{5}[-\i(\l_{\tau}+\q_{\tau})\cdot\bg_{\tau}+(\l+\q)\cdot\bg-m]\gamma_{5}[-\i\l_{\tau}\cdot\bg_{\tau}+\l\cdot\bg-m]\gamma_{j}\gamma_{5}}{[(\p_{\tau}-\l_{\tau})^{2}+c^{2}(\p-\l)^{2}+r][(\l_{\tau}+\q_{\tau})^{2}+(\l+\q)^{2}+m^{2}][\l_{\tau}^{2}+\l^{2}+m^{2}]}.\eea
This is rearranged as
\be\delta\Gamma_{5}\gamma_{5}=\lambda^{2}\int^{1}_{0}dxdydz\delta(1-x-y-z)\int\f{d^{d+\et}\tilde{l}}{(2\pi)^{d+\et}}\f{2\N}{\D^{3}},\ee
where $\D$, $\tilde{l}$, $\Delta$, and $a_{x}^{2}$ are given in Eq. (\ref{eq:deno}). The numerator is
\bea\N&=&\sum_{j}\gamma_{j}\gamma_{5}[-\i(\l_{\tau}+\q_{\tau})\cdot\bg_{\tau}+(\l+\q)\cdot\bg-m]\gamma_{5}[\l_{\tau}\cdot\bg_{\tau}+\l\cdot\bg-m]\gamma_{j}\gamma_{5}\non
&=&\sum_{j}\gamma_{j}\gamma_{5}[-\i\tilde{\l}_{\tau}\cdot\bg_{\tau}+\tilde{\l}\cdot\bg]\gamma_{5}[-\i\tilde{\l}_{\tau}\cdot\bg_{\tau}+\tilde{\l}\cdot\bg]\gamma_{j}\gamma_{5}+\O(\tilde{l})\non
&\rightarrow&-d[\tilde{\l}_{\tau}^{2}+\tilde{\l}^{2}]\gamma_{i}\gamma_{5}.\eea
In the last line, only the quadratic terms are kept. The integration is straightforward, and the result is
\be\delta\Gamma_{5}=-\lambda^{2}\int^{1}_{0}dxdydz\delta(1-x-y-z)\f{\Gamma(\f{4-d-\et}{2})[d+d^{2}/a_{x}^{2}]}{a_{x}^{d}2(4\pi)^{(d+1)/2}\Delta^{\f{4-d-\et}{2}}}.\ee
Near $d=3$ and $\et=1$, we find
\be\delta\Gamma_{5}=-\f{\lambda^{2}}{16\pi^{2}\epsilon}\int^{1}_{0}dxdydz\delta(1-x-y-z)(3a_{x}^{-3}+9a_{x}^{-5})=-\f{3\lambda^{2}}{4\pi^{2}\epsilon c(1+c)}.\label{re:vc6}\ee
This is the same with $\delta_{m}$ in Eq. (\ref{re:fermionself}), i.e., $Z_{5}=Z_{m}$. Thus, the Ward identity in Eq. (\ref{eq:wardidentity}) is proven.

\end{widetext}


\begin{thebibliography}{9}
\bibitem{DMS_Review_I} H. V. Lohneysen, \textit{Electron-electron interactions and the metal-insulator transition in heavily doped silicon}, Ann. Phys. (Berlin) {\textbf 523}, 599 (2011).
\bibitem{DMS_Review_II} V. Dobrosavljevic, \textit{Introduction to metal-insulator transitions, in Conductor-Insulator Quantum Phase Transitions}, edited by V. Dobrosavljevic, N. Trivedi, and J. M. Valles Jr. (Oxford University Press, Oxford, 2012).
\bibitem{DMS_Review_III} T. Dietl, and H. Ohno, \textit{Dilute ferromagnetic semiconductors: Physics and spintronic structures}, Rev. Mod. Phys. {\textbf 86}, 187 (2014).
\bibitem{Ref1} A. A. Burkov, M. D. Hook, and L. Balents, \textit{Topological nodal semimetals}, Phys. Rev. B \textbf{84}, 235126 (2011).
\bibitem{Ref2} C.-X. Liu, P. Ye, and X.-L.Qi, \textit{Chiral gauge field and axial anomaly in a Weyl semimetal} Phys. Rev. B \textbf{87}, 235306 (2013).
\bibitem{Ref3} G. Y. Cho, \textit{Possible topological phases of bulk magnetically doped $Bi_2Se_3$: turning a topological band insulator into Weyl semimetal}, arxiv:1110.1939 (2011).
\bibitem{Experiments_Doping_MI} Kyoung-Min Kim, Jinsu Kim, Soo-Whan Kim, Myung-Hwa Jung, and Ki-Seok Kim, \textit{Universal appearance of Weyl metals from magnetically doped semiconductors}, in preparation.
\bibitem{WM_Boltzmann_KSK} H.-J. Kim, K.-S. Kim, J.-F. Wang, M. Sasaki, N. Satoh, A. Ohnishi, M. Kitaura, M. Yang, and L. Li, \textit{Dirac versus Weyl fermions in topological insulators: Adler-Bell-Jackiw anomaly in transport phenomena}, Phys. Rev. Lett. \textbf{111}, 246603 (2013); K.-S. Kim, H.-J. Kim, and M. Sasaki, \textit{Boltzmann equation approach to anomalous transport in a Weyl metal}, Phys. Rev. B \textbf{89}, 195137 (2014); Ki-Seok Kim, Heon-Jung Kim, M. Sasaki, J.-F. Wang, and L. Li, \textit{Anomalous transport phenomena in Weyl metal beyond the Drude model for Landau's Fermi liquids}, Sci. Technol. Adv. Mater. \textbf{15}, 064401 (2014).
\bibitem{Nonlinear_LMR_WM} Dongwoo Shin, Yongwoo Lee, M. Sasaki, Yoon Hee Jeong, Franziska Weickert, Jon B. Betts, Heon-Jung Kim, Ki-Seok Kim, and Jeehoon Kim, \textit{Violation of Ohm's law in a Weyl metal}, Nature Materials \textbf{16}, 1096 (2017), doi:10.1038/nmat4965.
\bibitem{WMD_KKM_16} K.-M Kim, D.-W Shin, M. Sasaki, H.-J. Kim, J.-H. Kim, and K.-S. Kim, \textit{Two-parameter scaling theory of the longitudinal magnetoconductivity in a Weyl metal phase: Chiral anomaly, weak disorder, and finite temperature}, Phys. Rev. B \textbf{94}, 085128, (2016).
\bibitem{FeBiTe_EXP_KSK} H.-J. Kim, K.-S. Kim, J.-F. Wang, V. A. Kulbachinskii, K. Ogawa, M. Sasaki, A. Ohnishi, M. Kitaura, Y.-Y. Wu, L. Li, I. Yamamoto, J. Azuma, M. Kamada, and V. Dobrosavljevic, \textit{Topological phase transitions driven by magnetic phase transitions in $Fe_{x}Bi_{2}Te_{3}$ $(0\le x\le0.1)$ single crystals}, Phys. Rev. Lett. \textbf{110}, 136601 (2013).
\bibitem{DMTS_KKM_15} K.-M. Kim, Y.-S. Jho, and K.-S. Kim, \textit{Dilute magnetic topological semiconductor}, Phys. Rev. B \textbf{91}, 115125 (2015).
\bibitem{Anomaly_Chiral_Current} Iksu Jang, Jae-Ho Han, and Ki-Seok Kim, \textit{Anomalous Hall effects beyond Berry magnetic fields in a Weyl metal phase}, Phys. Rev. B \textbf{95}, 054117 (2017); Yong-Soo Jho, Jae-Ho Han, and Ki-Seok Kim, \textit{Topological Fermi-liquid theory for interacting Weyl metals with time reversal symmetry breaking}, Phys. Rev. B \textbf{95}, 205113 (2017); Iksu Jang and Ki-Seok Kim, \textit{Chiral pair of Fermi arcs, anomaly cancellation, and spin or valley Hall effects in Weyl metals with broken inversion symmetry}, Phys. Rev. B \textbf{97}, 165201 (2018).
\bibitem{TI_Reviews} M. Z. Hasan and C. L. Kane, \textit{Colloquium: Topological insulators}, Rev. Mod. Phys. {\textbf 82}, 3045 (2010); X.-L. Qi, and S.-C. Zhang, \textit{Topological insulators and superconductors}, Rev. Mod. Phys. {\textbf 83}, 1057 (2011).
\bibitem{RKKY} M. A. Ruderman and C. Kittel, \textit{Indirect exchange coupling of nuclear magnetic moments by conduction electrons}, Phys. Rev. \textbf{96}, 99 (1954); T. A. Kasuya, \textit{Theory of metallic ferro- and antiferromagnetism on Zener's model}, Prog. Theor. Phys. \textbf{16}, 45 (1956); K. Yosida, \textit{Magnetic properties of Cu-Mn alloys}, Phys. Rev. \textbf{106}, 893 (1957).
\bibitem{Phi4Theory_RG} J. Zinn-Justin, \textit{Quantum Field Theory and Critical Phenomena}, Clarendon Press 1989 (Oxford 4th ed. 2002).
\bibitem{double_epsilon} Boyanovsky, Daniel and Cardy, John L., \textit{Critical behavior of $m$-component magnets with correlated impurities}, Phys. Rev. B \textbf{26}, 154 (1982).
\bibitem{Peskin} M. Peskin and D. Schroeder, \textit{An Introduction to Quantum Field Theory}, Perseus Books, 1995.
\bibitem{Chiral_Anomaly_I} H. B. Nielsen, and M. Ninomiya, \textit{The Adler-Bell-Jackiw anomaly and Weyl fermions in a crystal}, Phys. Lett. B \textbf{130}, 389 (1983).
\bibitem{Chiral_Anomaly_II} F. D. M. Haldane, \textit{Berry curvature on the Fermi surface: Anomalous Hall effect as a topological Fermi-liquid property}, Phys. Rev. Lett. \textbf{93}, 206602 (2004).
\bibitem{Chiral_Anomaly_III} S. Murakami, \textit{Phase transition between the quantum spin Hall and insulator phases in 3D: Emergence of a topological gapless phase}, New J. Phys. \textbf{9}, 356 (2007).
\bibitem{Chiral_Anomaly_IV} A. A. Burkov and L. Balents, \textit{Weyl semimetal in a topological insulator multilayer}, Phys. Rev. Lett. \textbf{107}, 127205 (2011).
\bibitem{Haldane_Mapping} A. Auerbach, \textit{Interacting Electrons and Quantum Magnetism}, Springer, 1994.
\bibitem{Adler_Principle} S. L. Adler, \textit{Consistency Conditions on the Strong Interactions Implied by a Partially Conserved Axial-Vector Current}, Phys. Rev. \textbf{137}, B1022 (1965).
\bibitem{velocityRG} Bitan Roy, Vladimir Juricic, and Igor F. Herbut, \textit{Emergent Lorentz symmetry near fermionic quantum critical points in two and three dimensions}, J. High Energy Phys. 2016:18 (2016).
\end{thebibliography}
\end{document}